\documentclass[twocolumn]{aastex631}
\usepackage{multirow}
\usepackage{subfigure}
\usepackage{amsmath}
\usepackage{array}

\begin{document}

\title{From Nonparametric Distance Reconstruction to Testing the Etherington Relation and Cosmic Curvature Using 2D and 3D BAO Measurements}   

\author{Darshan Kumar}
\email{kumardarshan@hnas.ac.cn}   
\affiliation{Institute for Gravitational Wave Astronomy, Henan Academy of Sciences, Zhengzhou, 450046, China}

\author{Jie Zheng}
\affiliation{Institute for Gravitational Wave Astronomy, Henan Academy of Sciences, Zhengzhou, 450046, China}

\author{Zhi-Qiang You}
\affiliation{Institute for Gravitational Wave Astronomy, Henan Academy of Sciences, Zhengzhou, 450046, China}

\author{Da-Chun Qiang}
\affiliation{Institute for Gravitational Wave Astronomy, Henan Academy of Sciences, Zhengzhou, 450046, China}

\correspondingauthor{Jie Zheng}  
\email{zhengjie@mail.bnu.edu.cn}

\begin{abstract}
We present a joint test of cosmic curvature, $\Omega_{k0}$, and the cosmic distance-duality relation (CDDR) using the Etherington relation, which connects the luminosity and angular diameter distances at the same redshift. In this work, we combine the angular diameter distance measurements from recent Baryon Acoustic Oscillation (BAO) observations with luminosity distances reconstructed from Cosmic Chronometers data of Hubble parameter $H(z)$ using a non-parametric technique, Gaussian Process. A key part of our analysis is the systematic comparison of different BAO measurements (2D BAO, 3D BAO, and 3D DESI BAO) to determine whether any potential tension between angular and anisotropic BAO data affects constraints on the distance duality parameter $\eta(z)$ and $\Omega_{k0}$. We adopt four representative parameterizations of $\eta(z)$ to examine the correlation between $\eta(z)$ and $\Omega_{k0}$. Our results show no evidence for violation of the CDDR, with $\eta(z)$ consistent with unity at the 99\% confidence level for all BAO datasets and parameterizations. In all scenarios, the best-fit values of $\Omega_{k0}$ mildly favor a non-flat universe, although a spatially flat universe remains compatible at the 95\% confidence level. The constraints on $\eta_1$ and $\Omega_{k0}$ indicate slight variations across different BAO datasets, but the discrepancies between the 2D and 3D BAO measurements do not introduce any significant bias, and no statistically meaningful tension is observed. Our work provides robust constraints on cosmic curvature and the validity of the CDDR based on non-parametric distance reconstruction.
\end{abstract}


\section{Introduction}\label{sec_intr}
The cosmic distance duality relation (CDDR), also known as the Etherington relation \citep{1933PMag...15..761E}, is a cornerstone of modern observational cosmology. It connects the luminosity distance $(d_{L})$ and the angular diameter distance ($d_{A}$) directly, through the expression $d_{L}(z) = d_{A}(z)(1+z)^{2}$, where $z$ denotes the redshift. As originally proposed by Etherington, this identity follows three basic assumptions \citep{2007GReGr..39.1047E}: $i)$ spacetime is described by a metric theory of gravity, $ii)$ photons travel along unique null geodesics, and $iii)$ the number of photons is conserved. Under these conditions, one expects $\eta(z) = d_{L}(z)/\left(d_{A}(z)(1+z)^{2}\right) = 1$ at all redshifts. {If any of these assumptions is not satisfied, for example, due to photon number non-conservation, non-geodesic photon propagation, or departures from metric gravity, deviations from the standard CDDR may occur. In this case, any significant deviation $\eta(z) \neq 1$ can be interpreted as evidence for new physics, such as photon–axion conversions \citep{2004PhRvD..69j1305B,2004ApJ...607..661B}, cosmic opacity induced by intergalactic dust or exotic interactions \citep{2009ApJ...696.1727M,2012JCAP...12..028N}, or modifications to general relativity \citep{2004PhRvD..70h3533U,2017PhRvD..95f1501S,2021PhRvD.104h4079A}.} 

The current state of cosmology faces growing tensions between the early- and late-Universe measurements of cosmological parameters \citep{Riess2022,1807.06209,2012MNRAS.427..146H,2020NatAs...4..196D,2021PhRvD.103d1301H,2021A&A...646A.140H,2021ApJ...908L...9D,2022PhRvD.105b3520A}, especially the ``Hubble tension" \citep{2018AAS...23231902P,2020MNRAS.498.1420W}. 
Despite extensive efforts \citep{2016PhLB..761..242D,pantheon,2020arXiv200710716E,Riess2022}, these inconsistencies may point to new physics or to revisions of foundational assumptions, such as the validity of the CDDR.
As highlighted in Ref.~\citep{Renzi2023,Teixeira:2025czm,Alfano:2025gie}, a thorough investigation of the CDDR may provide valuable insights into pressing cosmological tensions, including the Hubble constant discrepancy and the possible redshift evolution of dark energy.
In parallel, the present-day spatial curvature $\Omega_{k0}$ could affect distance measurements through the transverse comoving distance. Consequently, violations of the CDDR could bias the measurement of spatial curvature. In addition, several works \citep{Bolejko2018,DiValentino2020NatAs,HeinesenBuchert2020,DiValentino2021APh,Handley2021PRD,Clifton2024} suggest that the inconsistency between the spatial curvature $\Omega_{k0}$ inferred from early- and late-Universe could impact the measurement of $H_0$. {Therefore, these considerations motivate us to explore the relation between the CDDR and the spatial curvature using a model-independent approach that does not assume a specific theoretical cosmological model for the expansion history, while relying on standard geometric relations.} 

A considerable amount of research has focused on testing the CDDR employing a variety of cosmological observations \citep{2010JCAP...10..024A,2010JCAP...02..008S,2012JCAP...06..022H,2013PhLB..718.1166L,2013JCAP...04..027H,2017IJMPD..2650097F,Kanodia:2025jqh,Zheng:2025cgq,2025ApJ...979....2Q}.
Beyond testing the CDDR alone, an increasing number of studies \citep{Qin2021MNRAS,Kumar2021PRD,2022JCAP...01..053K,Renzi2022MNRAS} has jointly determined the CDDR parameter $\eta(z)$ and the present-day spatial curvature $\Omega_{k0}$ using Type Ia supernovae (SNIa), baryon acoustic oscillations (BAO), cosmic chronometers (CC), high-redshift quasars, and strong gravitational lensing (SGL). These analyses consistently reveal a non-negligible $\eta$–$\Omega_{k0}$ degeneracy and find no significant violation of the CDDR. However, they found that the constraints are sensitive to the dataset selection, redshift coverage, and reconstruction methodology. 
{This work aims to extend these studies by systematically comparing different types of BAO measurements, including 2D BAO, 3D BAO, and 3D DESI BAO, within a unified analysis framework. A non-parametric reconstruction of the expansion history from cosmic chronometers is incorporated to explicitly assess how BAO measurements and $H_0$ priors influence the joint constraints on $\eta(z)$ and $\Omega_{k0}$. Hence, a renewed and minimally dependent on specific cosmological models joint analysis of $\eta(z)$ and $\Omega_{k0}$ is both timely and necessary.}

In addition, several additional considerations motivate a fresh, comprehensive investigation that constrains the CDDR parameter $\eta(z)$ and the present-day spatial curvature $\Omega_{k0}$ jointly. The Dark Energy Spectroscopic Instrument (DESI) collaboration released its second dataset (DR2, hereafter 3D–DESI), which is the most precise, anisotropic (3D) BAO measurements to date \citep{2025JCAP...04..012A}. {These data have sparked renewed discussion of potential deviations from $\Lambda$CDM and possible dynamical dark energy when combined with complementary probes such as the CMB and Type Ia supernovae \citep{2025JCAP...04..012A,2025JCAP...02..021A,Zheng:2024qzi,2024PDU....4601668H,Lu:2024hvv,2025ApJ...987...58W,Li:2025eqh,2025arXiv250502207Y,Abedin:2025dis,2025arXiv250524732C,Luetal2025}, prompting further investigations into testing the validity of CDDR and $\Omega_{k0}$ using the latest BAO measurements. }
Furthermore, several studies have reported the disagreements between BAO measurements obtained from the two-dimensional (2D, transverse or angular) BAO and the three-dimensional (3D, or anisotropic) BAO \citep{2019PhRvD..99l3515A,2020MNRAS.495.2630C,2023PhRvD.107j3531B,9rt7-ph33,2024Univ...10..406D,2024PhLB..85839027F,Zheng:2025cgq}. The inconsistencies between 2D and 3D BAO measurements would directly affect the constraints on $\Omega_{k0}$ and $\eta(z)$.  

In practice, a key challenge is to match the luminosity and angular diameter distances at the same redshifts in testing the CDDR. In our case, we reconstruct the luminosity distance–redshift relation from the Hubble parameter $H(z)$ using the Gaussian Process (GP) technique. 
The GP technique provides a non-parametric, model-independent reconstruction of the expansion history, which has been widely used in cosmology \citep{Liu2019ApJGPStrongLens,2021MNRAS.504.3938M,Zheng2021EPJC_DECIGO_GP_Omegak,WuQiZhang2023CPC_GPNullTest,Qi2023PRD_GPvANN_Curvature,Jesus2024MNRAS}. 
Compared to other reconstruction methods, such as the Artificial Neural Network (ANN) method, the GP technique offers several distinct advantages. {The GP technique naturally propagates observational uncertainties and their covariances into the reconstructed distance estimates and, due to its inherent regularization, produces smooth reconstruction with well-characterized uncertainties and is less prone to overfitting \citep{2012JCAP...06..036S,2012PhRvD..85l3530S}. These features make GP particularly suitable for our analysis, as it provides a robust reconstruction of luminosity distances without relying on a specific cosmological model.}

{In this work, we propose the joint test for the validity of the CDDR and constrain the cosmic curvature $\Omega_{k0}$ by combining luminosity distance $d_L$ and angular diameter distance $d_A$ measurements. For $d_L$, we reconstruct $H(z)$ from cosmic chronometer data using a non-parametric Gaussian Process method and compute the comoving distance $d_C$, from which $d_L$ is obtained. We consider three different $H_0$ priors to examine the effect of the Hubble constant on the inferred distances. For $d_A$, we consider 2D and 3D BAO measurements and check their systematic consistency. We then compare $d_L$ and $d_A$ through the distance duality relation parameter $\eta(z)$, adopting four different parameterizations to study the redshift evolution of the CDDR. }
This strategy not only enables a robust parameterized estimation of $\eta(z)$ and $\Omega_{k0}$, but also allows us to assess how the potential tensions between 2D and 3D BAO measurements would affect the inferred behavior of $\eta(z)$ and $\Omega_{k0}$. The outline of the paper is as follows: In section~\ref{sec_obse_data_meth}, we discuss the Data and Methodology. The analysis and results are explained in section~\ref{sec_resu}. Finally, the discussions and conclusions are presented in section~\ref{sec_disc_conc}.

\section{Observational Datasets and Methodology}\label{sec_obse_data_meth}   
In this section, we describe the observational datasets and the reconstruction method we used. {Here, we consider two types of BAO measurements: $(i)$ the 2D (angular) BAO measurements including 15 measurements of \(\theta_{\mathrm{BAO}}\) (see Table~\ref{tab:2D}); and $(ii)$ the 3D (anisotropic) BAO measurements reported as \(d_A(z)/r_d\) (see Table~\ref{tab:3D}).}
For the luminosity distance $d_L(z)$, we reconstruct $H(z)$ from Cosmic Chronometers in a non-parametric approach using Gaussian Processes (GP).
Then, we combine $d_A(z)$ and $d_L(z)$ to constrain $\eta(z)$ and the spatial curvature parameter, $\Omega_{k0}$.
{
In this work, we consider four parameterizations of the cosmic distance–duality relation:

\begin{itemize}
    \item \textbf{P1 (Linear):} $\eta(z) = 1 + \eta_1 \times z$
    \item \textbf{P2 (Modified linear):} $\eta(z) = 1 + \eta_1 \times  \dfrac{z}{1 + z}$
    \item \textbf{P3 (Logarithmic):} $\eta(z) = 1 + \eta_1  \ln(1 + z)$
    \item \textbf{P4 (Power-law):} $\eta(z) = (1 + z)^{\eta_1}$
\end{itemize}
In all cases, setting $\eta_1 = 0$ recovers the standard form of the CDDR.


}
{In the following subsections, we provide further details on the datasets and reconstruction methodology.}

\subsection{The 2D and 3D BAO datasets}  
The clustering of matter preserves a characteristic comoving scale set by the sound horizon at the baryon drag epoch of the early Universe. This scale is imprinted on the matter distribution by acoustic oscillations in the early photon–baryon plasma. As the Universe expands, it is stretched to a present-day comoving galaxy separation of $r_d \simeq 150\,\mathrm{Mpc}$, {$r_d$ denotes the sound horizon at the drag epoch} and serves as a standard ruler in cosmology. The definition of $r_d$ is given by   
\begin{equation}
\label{eq:rdequation}
r_d=\int_{z_d}^{\infty}\frac{c_s(z)}{H(z)}\,dz,
\end{equation}
with $z_d$ is the redshift of the drag epoch, \(c_s(z)\) is the sound speed and \(H(z)\) is the Hubble parameter. Then, the angular BAO scale, \(\theta_{\mathrm{BAO}}\), is defined as
\begin{equation}
\label{eq:theta}
\theta_{\mathrm{BAO}}=\frac{r_d}{(1+z)\,d_A(z)},
\end{equation}
where \(d_A(z)\) is the angular-diameter distance and the comoving distance is \(d_M(z)=(1+z)d_A(z)\). 
In our work, we use three BAO datasets that provide $d_A(z)$ information, including 2D BAO, 3D BAO, and 3D DESI BAO. The 2D BAO measurements are derived from SDSS DR7, DR10, DR11, DR12, and DR12Q \citep{2020MNRAS.497.2133N,2021A&A...649A..20D,2016PhRvD..93b3530C,Alcaniz:2016ryy,2020APh...11902432C,2018JCAP...04..064D} and are obtained without adopting a fiducial cosmological model. {For the 3D BAO datasets, we consider results from DES Y6 and BOSS/eBOSS \citep{2017MNRAS.465.1757G,DES:2024pwq,2021MNRAS.500.1201H,2020ApJ...901..153D,2017MNRAS.470.2617A,2021PhRvD.103h3533A}, as well as the recent DESI DR2 release, hereafter denoted as 3D DESI BAO \citep{2025JCAP...04..012A}}. To remain model– and calibrator–independent, we use only the angular components of the 3D BAO measurements and exclude the radial and dilation scale data. It is crucial to emphasize that cosmological inferences from BAO measurements sensitively depend on the adopted $r_d$. {In this work, we treat the drag epoch sound horizon $r_d$ as a free parameter and marginalize it with a wide flat prior $\mathcal{U}[0,200]$}. 
    
\begin{table*}[htbp]
\renewcommand{\arraystretch}{1.4}                                          
\centering
\begin{tabular}{lcccc}
\hline
Survey & $z$ & $\theta_{\mathrm{BAO}}$ [deg] & References \\
\hline
SDSS DR12 & 0.11 & $19.8 \pm 3.26$ & de Carvalho et al. (2021) \\
\hline
\multirow{2}{*}{SDSS DR7} 
          & 0.235 & $9.06 \pm 0.23$ & \multirow{2}{*}{Alcaniz et al. (2017)} \\ 
\cline{2-3}
          & 0.365 & $6.33 \pm 0.22$ & \\
\hline
\multirow{6}{*}{SDSS DR10} 
          & 0.45  & $4.77 \pm 0.17$ & \multirow{6}{*}{Carvalho et al. (2016)} \\
\cline{2-3}
          & 0.47  & $5.02 \pm 0.25$ & \\
\cline{2-3}
          & 0.49  & $4.99 \pm 0.21$ & \\
\cline{2-3}
          & 0.51  & $4.81 \pm 0.17$ & \\
\cline{2-3}
          & 0.53  & $4.29 \pm 0.30$ & \\
\cline{2-3}
          & 0.55  & $4.25 \pm 0.25$ & \\
\hline
\multirow{5}{*}{SDSS DR11} 
          & 0.57  & $4.59 \pm 0.36$ & \multirow{5}{*}{Carvalho et al. (2020)} \\
\cline{2-3}
          & 0.59  & $4.39 \pm 0.33$ & \\
\cline{2-3}
          & 0.61  & $3.85 \pm 0.31$ & \\
\cline{2-3}
          & 0.63  & $3.90 \pm 0.43$ & \\
\cline{2-3}
          & 0.65  & $3.55 \pm 0.16$ & \\
\hline
BOSS DR12Q & 2.225 & $1.77 \pm 0.31$ & de Carvalho et al. (2018) \\
\hline
\end{tabular}
\caption{List of the 15 2D BAO data points used in this work, with $\theta_{\mathrm{BAO}}(z)\,[\mathrm{rad}] = r_d / [(1 + z)d_A(z)]$. The values in the third column are given in degrees. {See the quoted references for details \citep{2021A&A...649A..20D,Alcaniz:2016ryy,2016PhRvD..93b3530C,2020APh...11902432C,2018JCAP...04..064D}.}}
\label{tab:2D}
\end{table*}

\begin{table*}[htbp]
\renewcommand{\arraystretch}{1.4}

\centering
\begin{tabular}{lcccc}
\hline
Survey & $z$ & $d_A(z)/r_d$ & References \\
\hline
\multirow{2}{*}{BOSS DR12}
        & 0.32 & $6.5986 \pm 0.1337$ & \multirow{2}{*}{Gil-Mar\'in et al.\ (2017)} \\
\cline{2-3}
        & 0.57 & $9.389 \pm 0.103$ & \\
\hline
DES Y6 & 0.85 & $2.932 \pm 0.068$ & Abbott et al. (2024) \\
\hline
eBOSS DR16Q & 1.48 & $12.18 \pm 0.32$ & Hou et al.\ (2021) \\  
\hline
eBOSS DR16 Ly$\alpha$-F & 2.334 & $11.25^{+0.36}_{-0.33}$ & du Mas des Bourboux et al.\ (2020) \\
\hline
{DESI DR2 LRG1} & 0.510  & $8.998 \pm 0.112$  & \multirow{6}{*}{Abdul Karim et al. (2025)} \\
\cline{1-3}
   {DESI DR2 LRG2}     & 0.706  & $10.168 \pm 0.106$ & \\
\cline{1-3}
   {DESI DR2 LRG3+ELG1}     & 0.934  & $11.155 \pm 0.080$ & \\
\cline{1-3}
   {DESI DR2 ELG2}     & 1.321  & $11.894 \pm 0.138$ & \\
\cline{1-3}
   {DESI DR2 QSO}     & 1.484  & $12.286 \pm 0.305$ & \\
\cline{1-3}
    {DESI DR2 Ly$\alpha$}    & 2.330  & $11.708 \pm 0.159$ & \\
\hline
\end{tabular}
\caption{Summary of 3D BAO measurements used in this work, with $d_A(z)/r_d$. {See the quoted references for details \citep{2017MNRAS.465.1757G,DES:2024pwq,2021MNRAS.500.1201H,2020ApJ...901..153D,2025PhRvD.112h3515A}.} In our analysis, we adopt two alternative 3D BAO datasets: the first five rows correspond to the BOSS/eBOSS data points, while the remaining rows include measurements from the DESI DR2.} 
\label{tab:3D}    
\end{table*}

\subsection{The Cosmic Chronometers}
The Hubble parameter $ H(z) $ is a fundamental quantity in cosmology that describes the expansion history of the universe. A direct method to measure this parameter uses the differential ages of passively evolving galaxies, an approach known as the ``Cosmic Chronometers" (CC) \citep{2003ApJ...593..622J,2005PhRvD..71l3001S}.

The Hubble parameter in terms of redshift is defined as:
\begin{equation}
    H(z) = -\frac{1}{1+z} \frac{dz}{dt}.
\end{equation}

To determine $ H(z) $, one must measure the small change in cosmic time $ dt $ that corresponds to a small change in redshift $ dz $. This requires the observation of galaxies that have evolved passively. In young, star-forming galaxies, continuous star birth means their light comes mainly from new stars. In contrast, passively evolving red galaxies are dominated by old stars, which makes them suitable for measuring the universe's differential aging \citep{2003ApJ...593..622J}.

For this method, astronomers select early-type, passively evolving galaxies with similar metallicity within a very small redshift range. The redshift difference $ \Delta z $ is found through precise spectroscopic observations. To estimate the age difference $ \Delta t $, researchers use a direct spectroscopic feature, the 4000~\AA~ break, which has a linear relation with the age of a galaxy's stellar population at a fixed metallicity \citep{2012JCAP...08..006M}.

{The dominant sources of systematic uncertainty in the Cosmic Chronometers method arise from stellar population synthesis modeling, assumptions about metallicity, and possible residual star formation. These effects are mitigated through the selection of massive, passively evolving galaxies within narrow redshift and metallicity bins, and they primarily affect the overall normalization rather than the redshift evolution of $H(z)$. Consequently, they are acceptable for the present analysis.}

Since this measurement of $ H(z) $ relies only on spectroscopic data, it does not depend on any specific cosmological model. This independence makes Cosmic Chronometers a strong tool for testing and constraining cosmological models. The compilation of CC data used in this work includes 32 measurements of $ H(z) $ across a redshift range of $ 0.07 \leq z \leq 1.965 $ \citep{2018MNRAS.476.1036M,2022ApJ...928L...4B,2023IJMPD..3250039K}.

\subsection{Gaussian Process Reconstruction}

To compute the angular diameter distance without a cosmological model, we first find the comoving distance $d_{co}$ from the Cosmic Chronometer $H(z)$ data. The relation is

\begin{equation}
    d_{co} = c\displaystyle\int_0^z \frac{dz'}{H(z')},
    \label{eq:dcdh}
\end{equation} 

A direct numerical integration of Eq. \ref{eq:dcdh} requires a continuous function for $E(z)$. Since the data consist of discrete $H(z)$ measurements, we use Gaussian Process (GP) regression to reconstruct a smooth $H(z)$ function \citep{2006gpml}. GP is a non-parametric method that places a probability distribution over possible functions that fit the data.

The GP is defined by a mean function and a covariance function. The covariance function $K(z, z')$ describes the correlation between function values at different redshifts:
\begin{equation}
    K(z, z') = \langle (H(z)-\mu(z))(H(z')-\mu(z')) \rangle.
\end{equation}

We set the prior mean function $\mu(z)$ to zero, which is a common choice as the data can be normalized to have zero mean. Tests with different mean functions showed that the results do not depend on this choice.

The method assumes each measurement comes from an independent Gaussian distribution from the same population, and that the correlation between measurements at two redshifts depends on their separation. We use a square exponential kernel: 
\begin{equation}
    K(z,z')= \sigma_f^2 \exp\left(-\frac{(z-z')^2}{2\ell^2}\right),
\end{equation}
where $\sigma_f$ and $\ell$ are hyperparameters that set the amplitude and correlation length. These parameters are found by maximizing the marginal log-likelihood of the data with flat priors.

We checked the sensitivity of our results to the kernel choice by repeating the analysis with Matérn class kernels (5/2, 7/2, 9/2). {The reconstructed Hubble distances from all kernels are in excellent agreement with those obtained using the squared exponential kernel, with $1\sigma$ confidence regions largely overlapping and differences in reconstructed distances typically less than ~1–2\%. Therefore, our main conclusions remain unchanged, and for consistency, we used the squared exponential kernel in this work.}


Using the reconstructed $H(z)$, we numerically integrate Eq.~\ref{eq:dcdh} with the Simpson 3/8 rule to obtain the comoving distance $d_{\mathrm{co}}$ at the redshifts corresponding to the BAO samples.

{The luminosity distance is related to the comoving distance within the standard FLRW framework as
\begin{equation}
  d_L(z) =
  \begin{cases}
    \dfrac{d_H(1+z)}{\sqrt{\Omega_{k0}}}
    \sinh\!\left[\sqrt{\Omega_{k0}}\,\dfrac{d_{\mathrm{co}}}{d_H}\right],
    & \mbox{for $\Omega_{k0} > 0$}, \\[8pt]
    d_{\mathrm{co}}(1+z),
    & \mbox{for $\Omega_{k0} = 0$}, \\[8pt]
    \dfrac{d_H(1+z)}{\sqrt{|\Omega_{k0}|}}
    \sin\!\left[\sqrt{|\Omega_{k0}|}\,\dfrac{d_{\mathrm{co}}}{d_H}\right],
    & \mbox{for $\Omega_{k0} < 0$}.
  \end{cases}
  \label{rdef}
\end{equation}

Here, $d_H = c/H_0$ denotes the Hubble distance, where $c$ is the speed of light and $H_0$ is the Hubble constant. The luminosity distance $d_L$ is computed from the reconstructed comoving distance using Eq.~\ref{rdef}, and its associated uncertainty $\sigma_{d_L}$ is obtained by propagating the errors from the Gaussian Process reconstruction of $d_{\mathrm{co}}(z)$. We note that $H(z)$ is reconstructed non-parametrically using Gaussian Processes, while the relation between comoving, luminosity, and angular diameter distances is evaluated within the standard FLRW framework, assuming a homogeneous and isotropic universe with constant spatial curvature.} 

In this analysis, we consider three different choices for the Hubble constant $H_0$ when estimating the luminosity distance:

\begin{itemize}
    \item \textbf{No prior:} {$H_0$ is directly obtained from the Gaussian Process reconstruction of $H(z)$ using Cosmic Chronometers: $H_0 = 64.87 \pm 4.47~\mathrm{km~s^{-1}~Mpc^{-1}}$.}
    \item \textbf{Planck prior:} $H_0 = 67.66 \pm 0.42~\mathrm{km~s^{-1}~Mpc^{-1}}$ \citep{1807.06209}.   
    \item \textbf{SH0ES prior:} $H_0 = 73.04 \pm 1.04~\mathrm{km~s^{-1}~Mpc^{-1}}$ \citep{2022ApJ...934L...7R}. 
\end{itemize}

This allows us to check the sensitivity of our constraints on the CDDR parameter $\eta$ and the curvature parameter $\Omega_{k0}$ to the choice of $H_0$ prior, and to verify the robustness of our results.      

For the parameter estimation, we focus on two quantities: the cosmic distance duality parameter $\eta_1$ and the cosmic curvature parameter $\Omega_{k0}$. To minimize the influence of priors on the results, we adopt broad uniform priors, $\mathcal{U}[-1,1]$ for $\eta_1$ and $\mathcal{U}[-2,2]$ for $\Omega_{k0}$. 
{To estimate the parameters, we apply Bayesian inference using the Python module \textbf{\texttt{emcee}}\footnote{\url{https://emcee.readthedocs.io/en/stable/}}, an affine-invariant Markov chain Monte Carlo (MCMC) sampler \cite{emcee}. The MCMC analysis uses 20 walkers with 10,000 steps per walker. This setup allows a thorough exploration of the parameter space. We discard the first 30\% of samples from each chain as burn-in to remove any bias from the initial conditions. The remaining samples are used to construct the posterior distributions. We test the convergence of the chains through two methods. First, we inspect the trace plots for each parameter to confirm proper mixing and stability around the best-fit values. Second, we compute the integrated auto-correlation time $\tau_f$ using the \textbf{\textit{autocorr.integrated\_time}} function from \textbf{\textit{emcee}}. This procedure accounts for the correlated nature of ensemble samplers and provides statistically reliable estimates of the posterior distributions.}


\section{Results and Discussion}\label{sec_resu}

\begin{figure*}[htbp]
\centering
\subfigure[P1: $\eta(z)$=1+$\eta_1\times z$]{\includegraphics[width=0.45\textwidth]{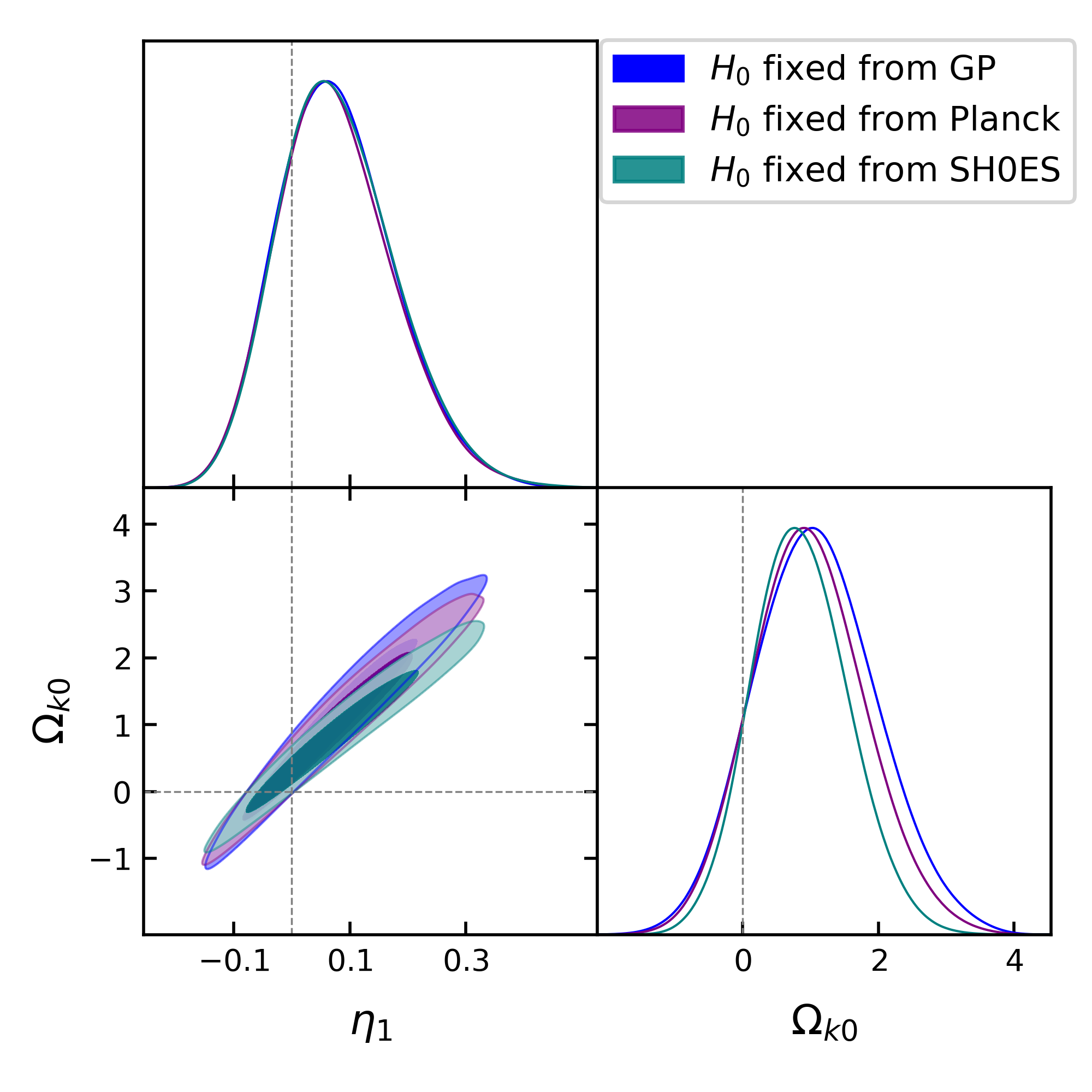}}\hspace{0.3cm}
\subfigure[P2: $\eta(z)$=1+$\eta_1\times \dfrac{z}{1+z}$]{\includegraphics[width=0.45\textwidth]{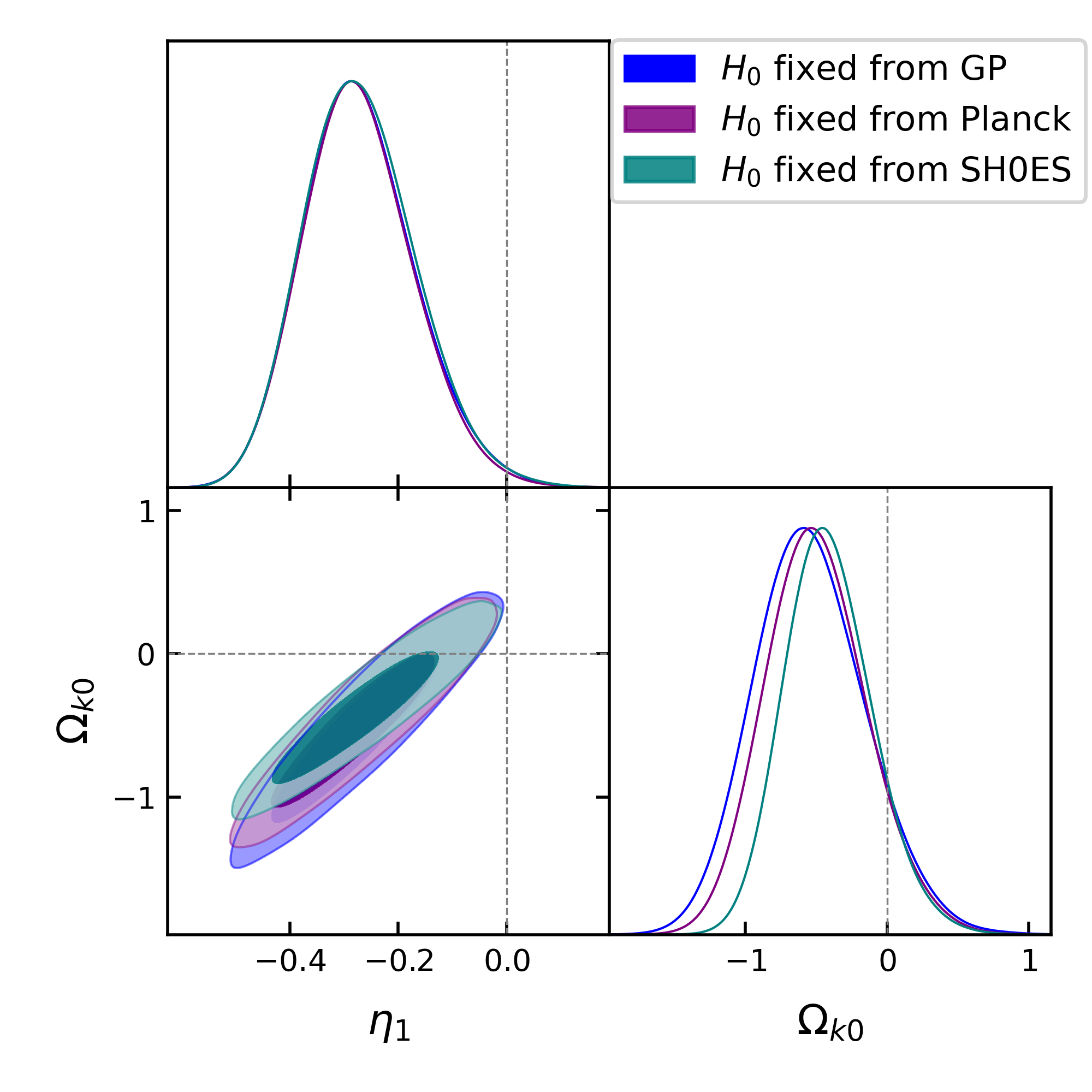}}\\
\subfigure[P3: $\eta(z)=1+\eta_1\ln{(1+z)}$]{\includegraphics[width=0.45\textwidth]{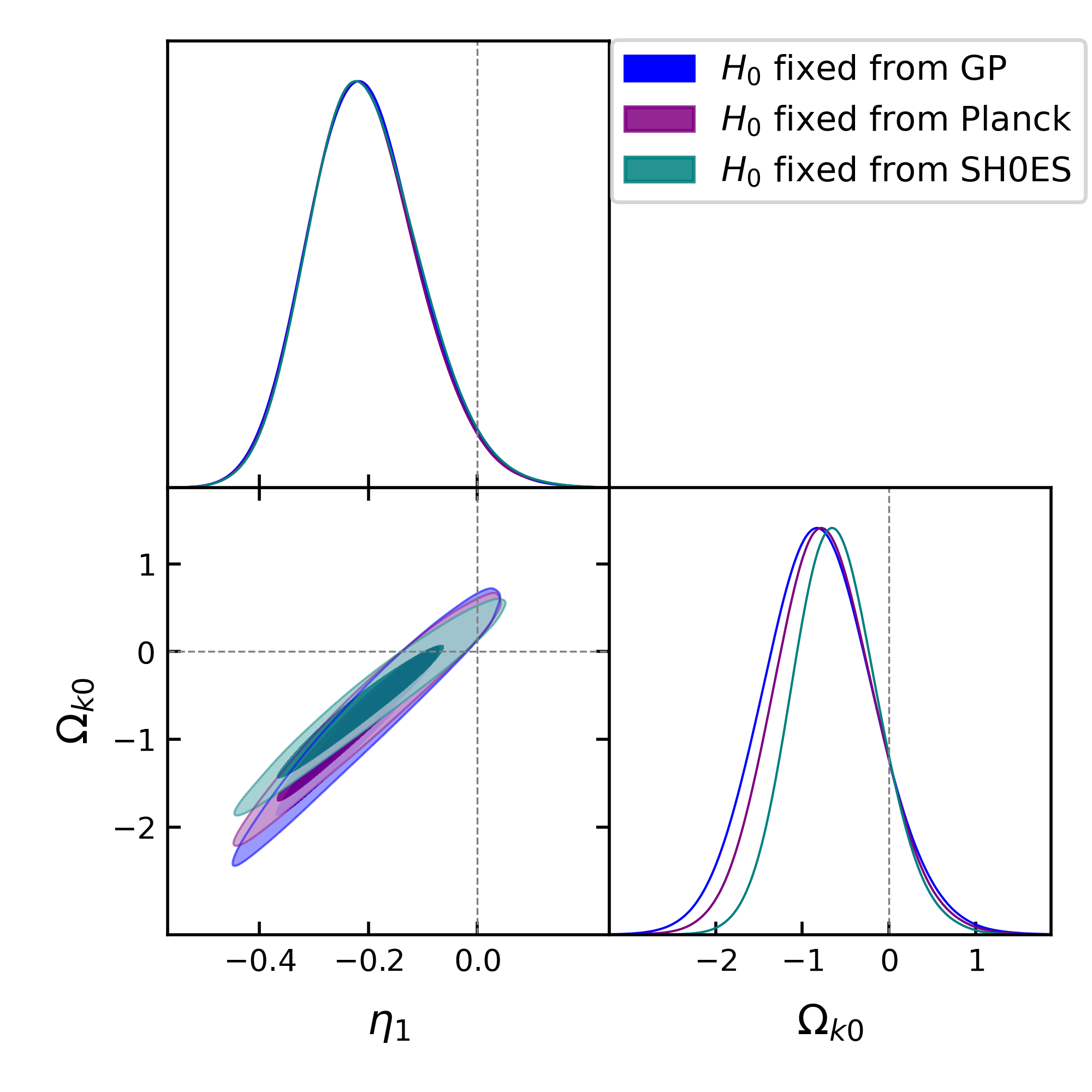}}\hspace{0.3cm}
\subfigure[P4: $\eta(z)={(1+z)}^{\eta_1}$]{\includegraphics[width=0.45\textwidth]{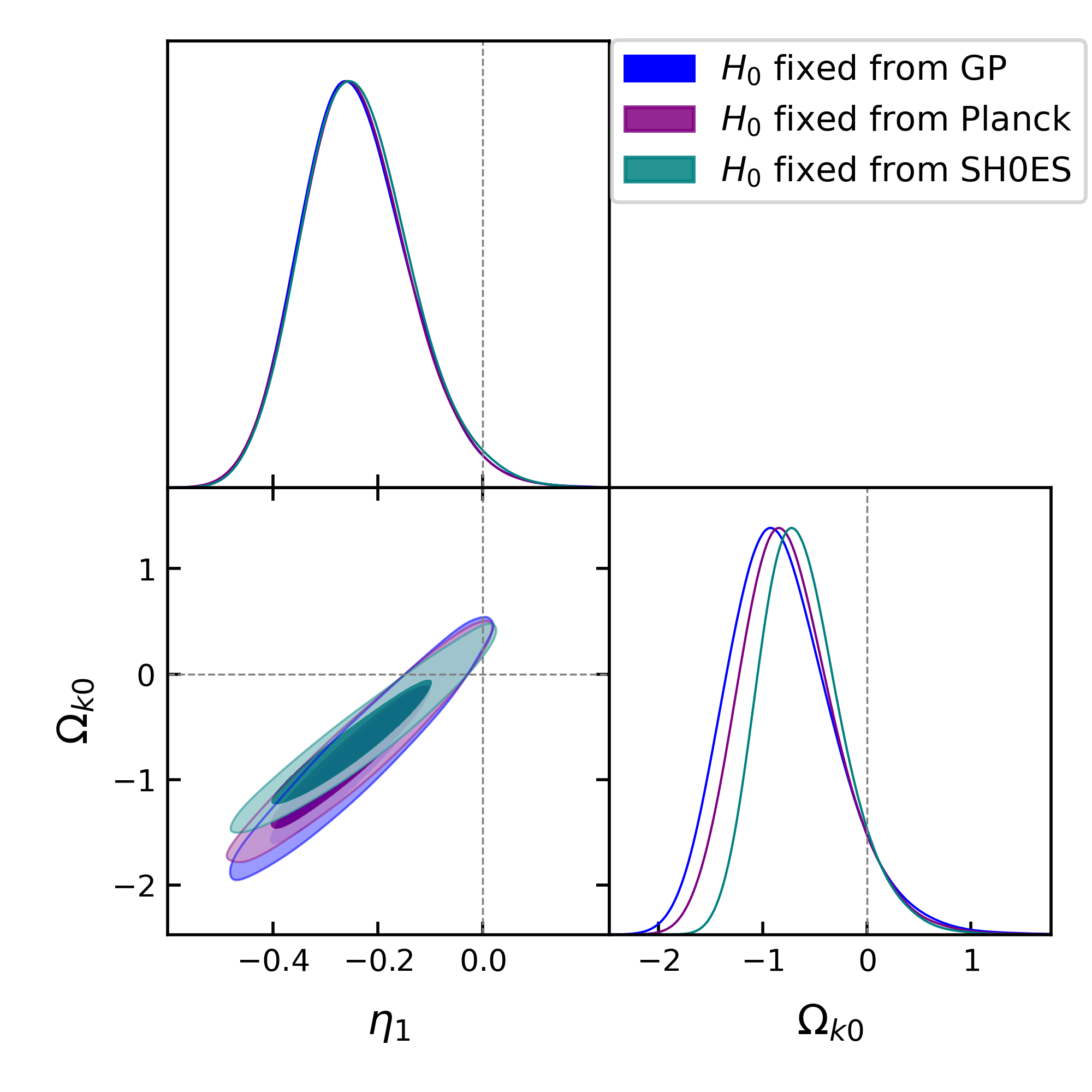}}
\caption{Posterior distributions and 68\% and 95\% confidence levels for $\eta_1$ and $\Omega_{k0}$ corresponding to the parameterizations P1–P4. The results use 2D BAO measurements with different $H_0$ priors (GP, Planck, SH0ES), in direct correspondence with Table~\ref{tab_cddr_bao_2d_results}. The dashed vertical lines indicate $\eta_1 = 0$ and $\Omega_{k0} = 0$, representing the standard cosmic distance duality relation and a spatially flat universe, respectively, to highlight potential deviations from these standard values.}
\label{fig_cddr_bao_2d_contours}
\end{figure*}

\begin{figure*}[htbp]
\centering       
\subfigure[P1: $\eta(z)$=1+$\eta_1\times z$]{\includegraphics[width=0.45\textwidth]{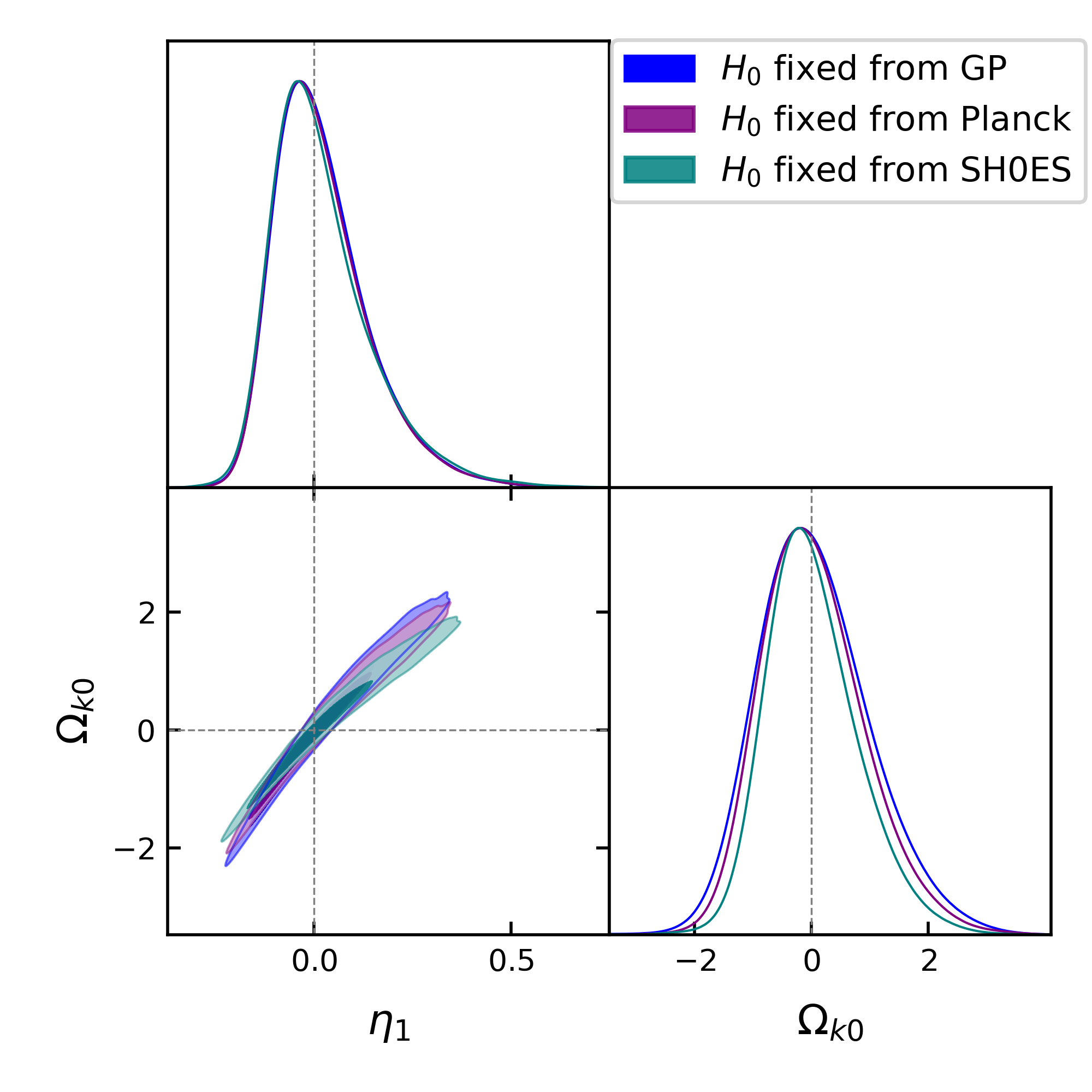}}\hspace{0.3cm}
\subfigure[P2: $\eta(z)$=1+$\eta_1\times \dfrac{z}{1+z}$]{\includegraphics[width=0.45\textwidth]{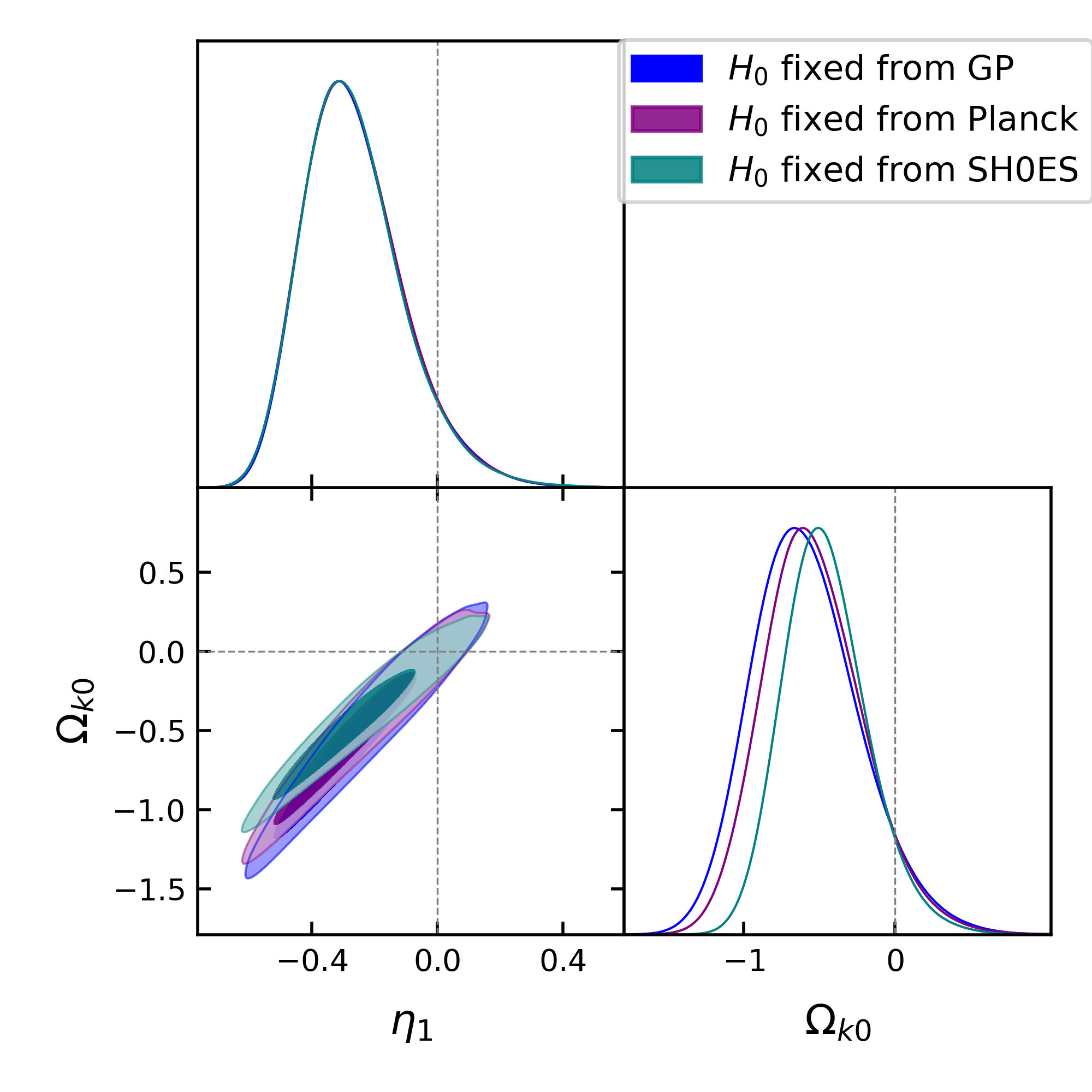}}\\
\subfigure[P3: $\eta(z)=1+\eta_1\ln{(1+z)}$]{\includegraphics[width=0.45\textwidth]{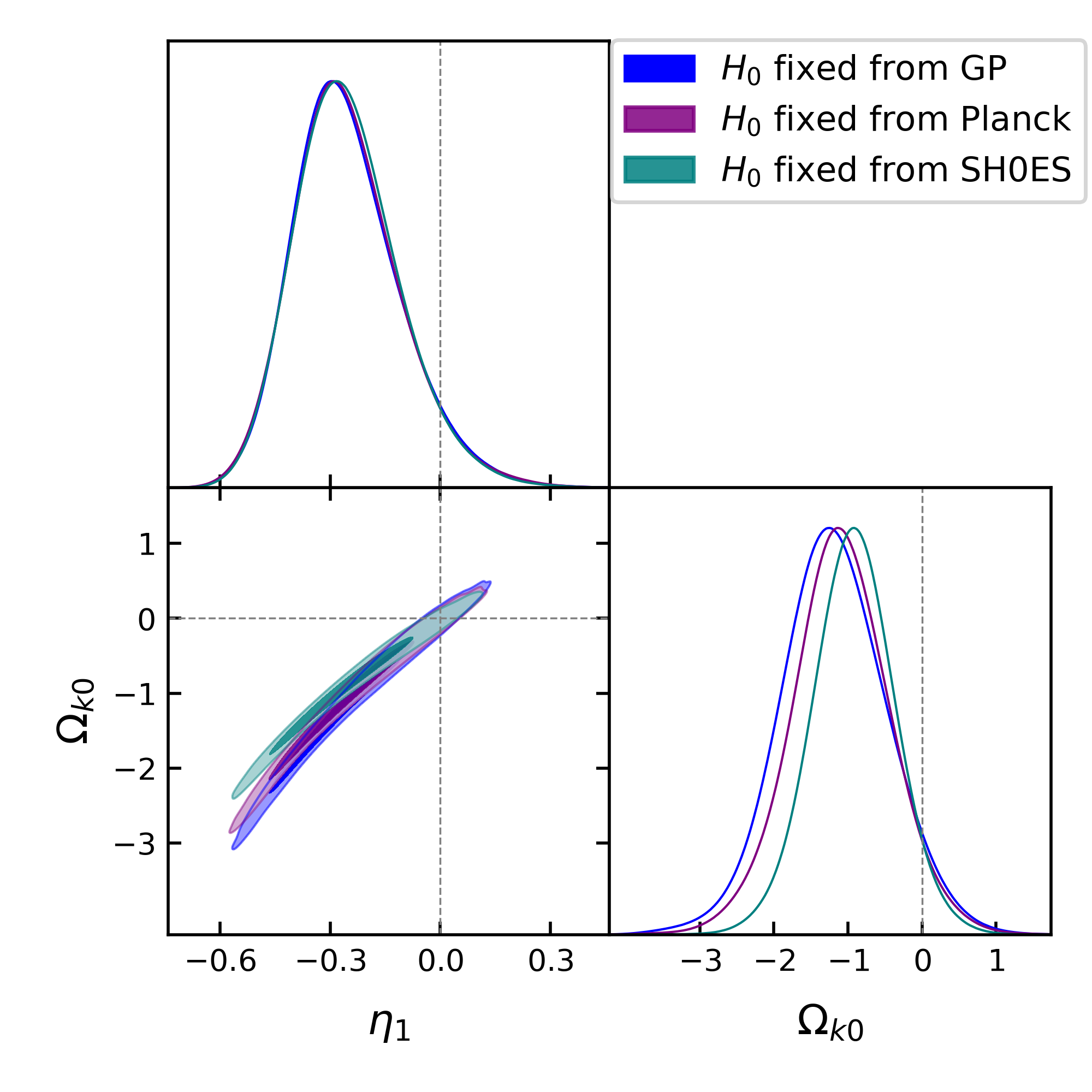}}\hspace{0.3cm}
\subfigure[P4: $\eta(z)={(1+z)}^{\eta_1}$]{\includegraphics[width=0.45\textwidth]{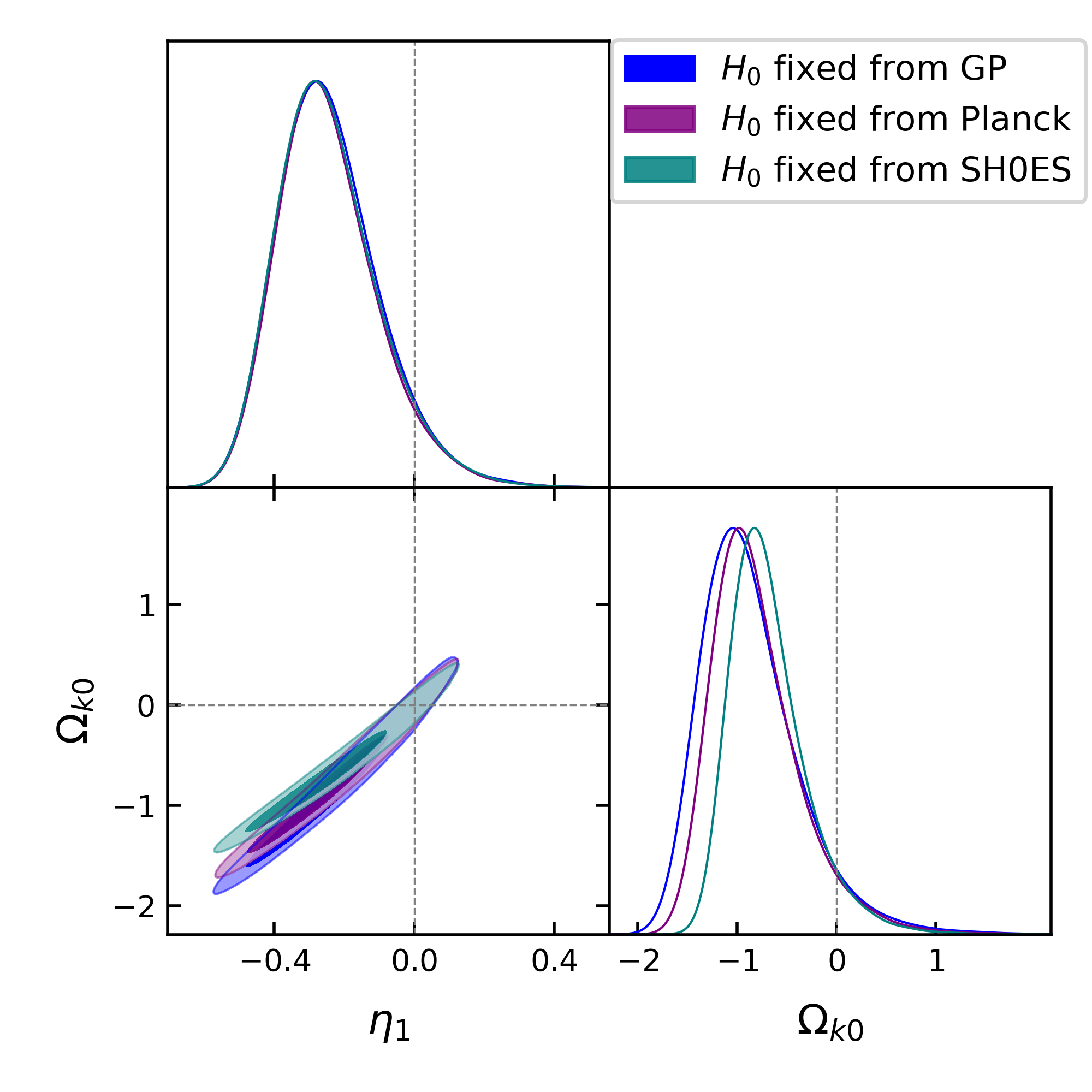}}
\caption{Posterior distributions and 68\% and 95\% confidence levels for $\eta_1$ and $\Omega_{k0}$ corresponding to the parameterizations P1–P4. The results use 3D BAO measurements with different $H_0$ priors (GP, Planck, SH0ES), in direct correspondence with Table~\ref{tab_cddr_bao_3d_results}. The dashed vertical lines indicate $\eta_1 = 0$ and $\Omega_{k0} = 0$, representing the standard cosmic distance duality relation and a spatially flat universe, respectively, to highlight potential deviations from these standard values.}
\label{fig_cddr_bao_3d_contours}    
\end{figure*}

\begin{figure*}[htbp]
\centering
\subfigure[P1: $\eta(z)$=1+$\eta_1\times z$]{\includegraphics[width=0.45\textwidth]{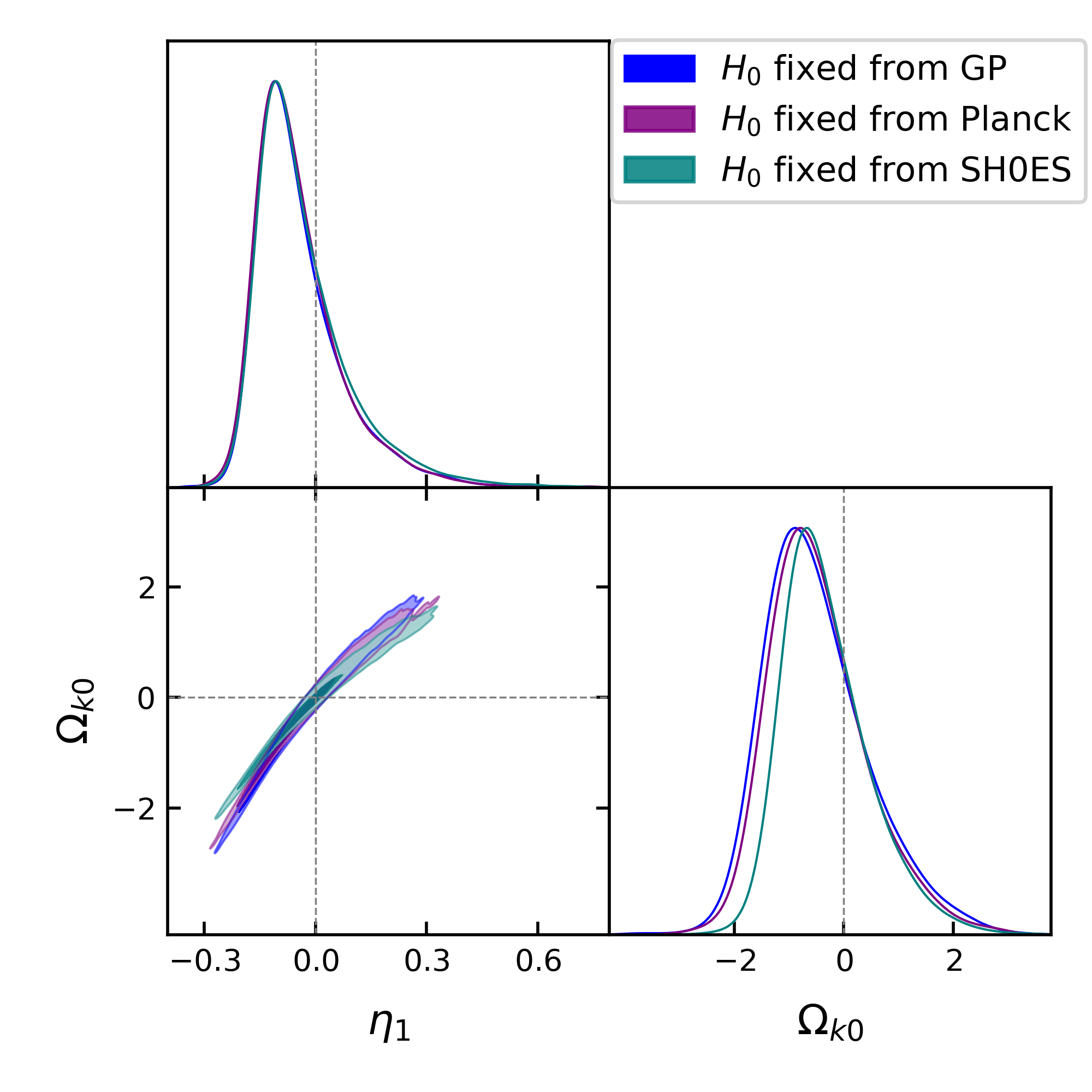}}\hspace{0.3cm}
\subfigure[P2: $\eta(z)$=1+$\eta_1\times \dfrac{z}{1+z}$]{\includegraphics[width=0.45\textwidth]{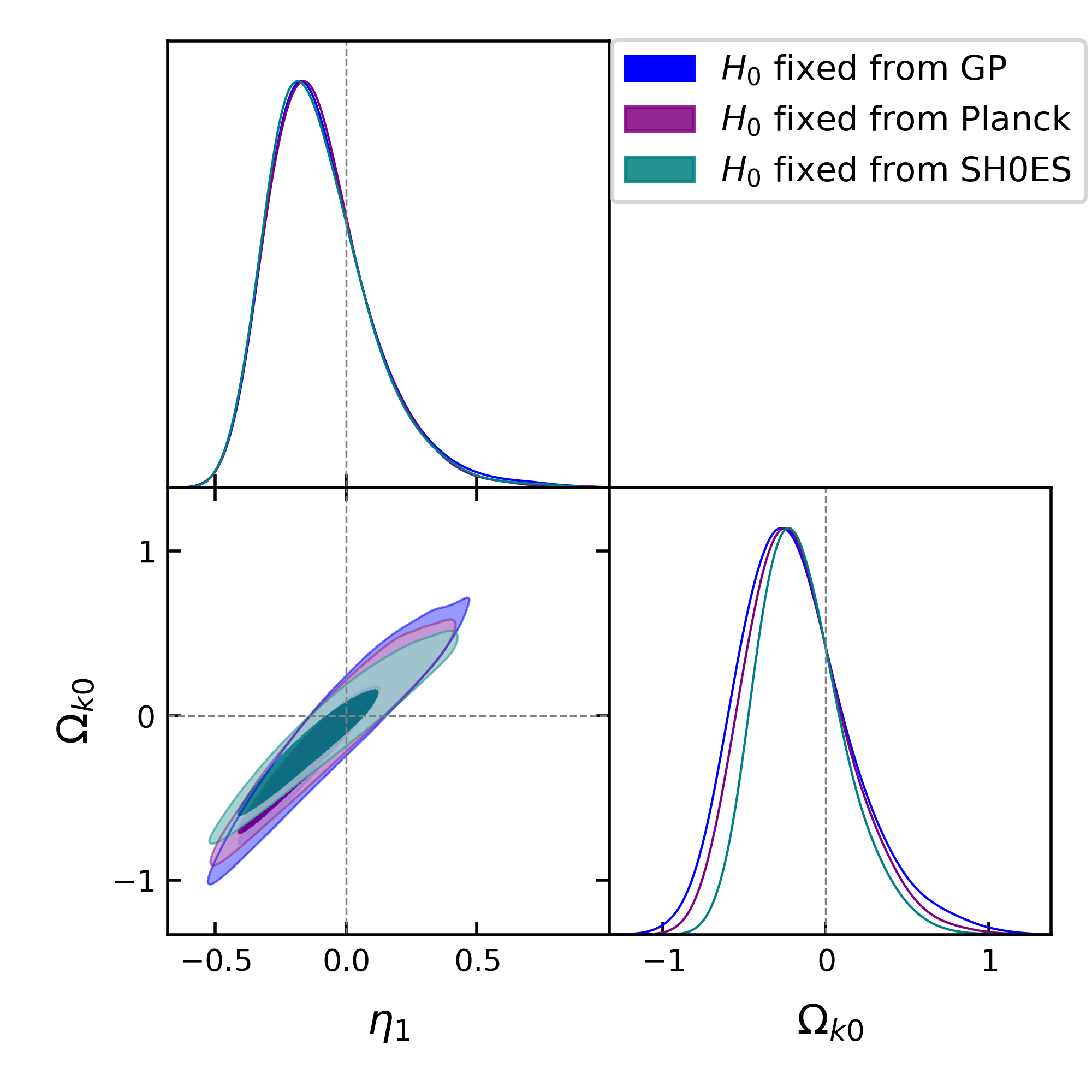}}\\
\subfigure[P3: $\eta(z)=1+\eta_1\ln{(1+z)}$]{\includegraphics[width=0.45\textwidth]{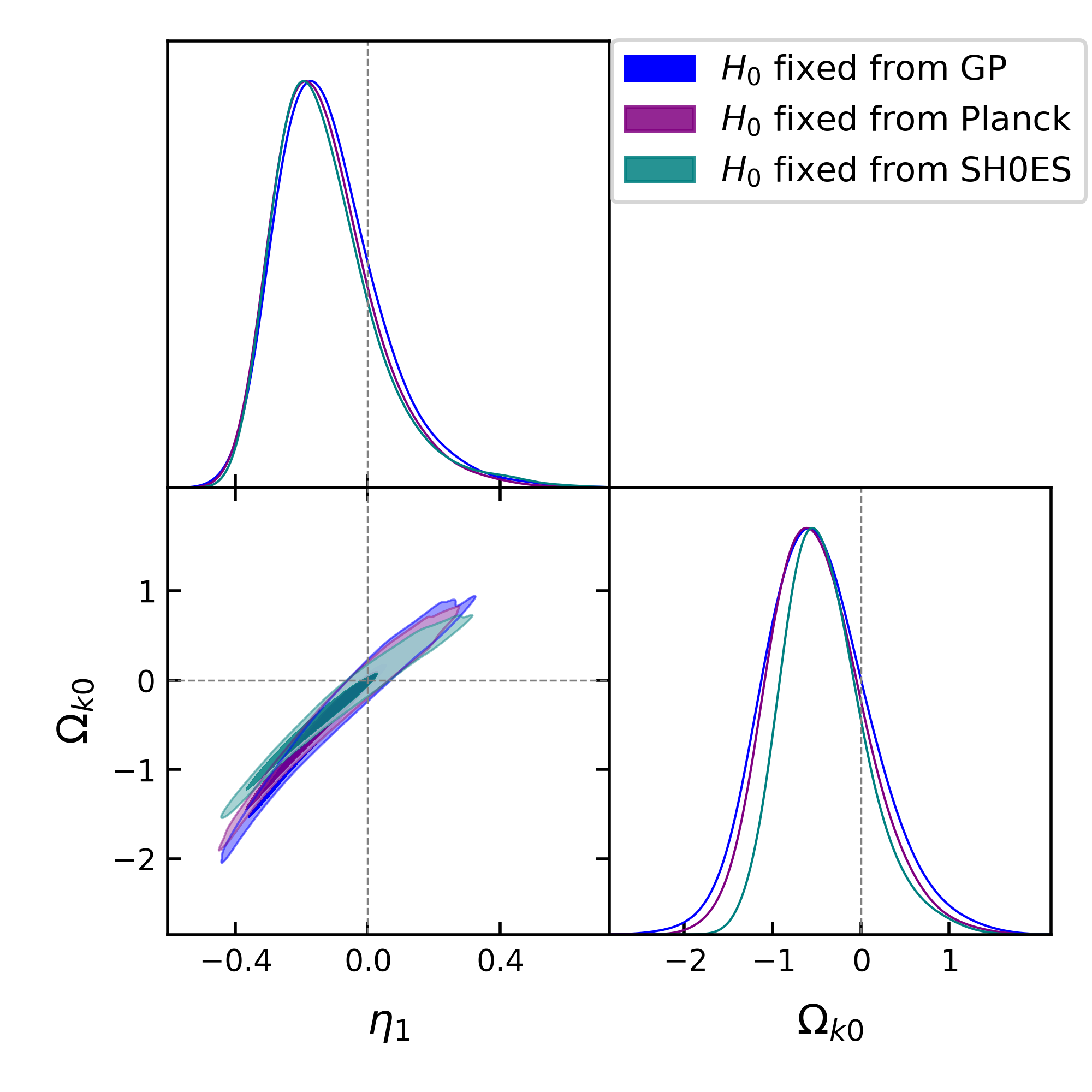}}\hspace{0.3cm}
\subfigure[P4: $\eta(z)={(1+z)}^{\eta_1}$]{\includegraphics[width=0.45\textwidth]{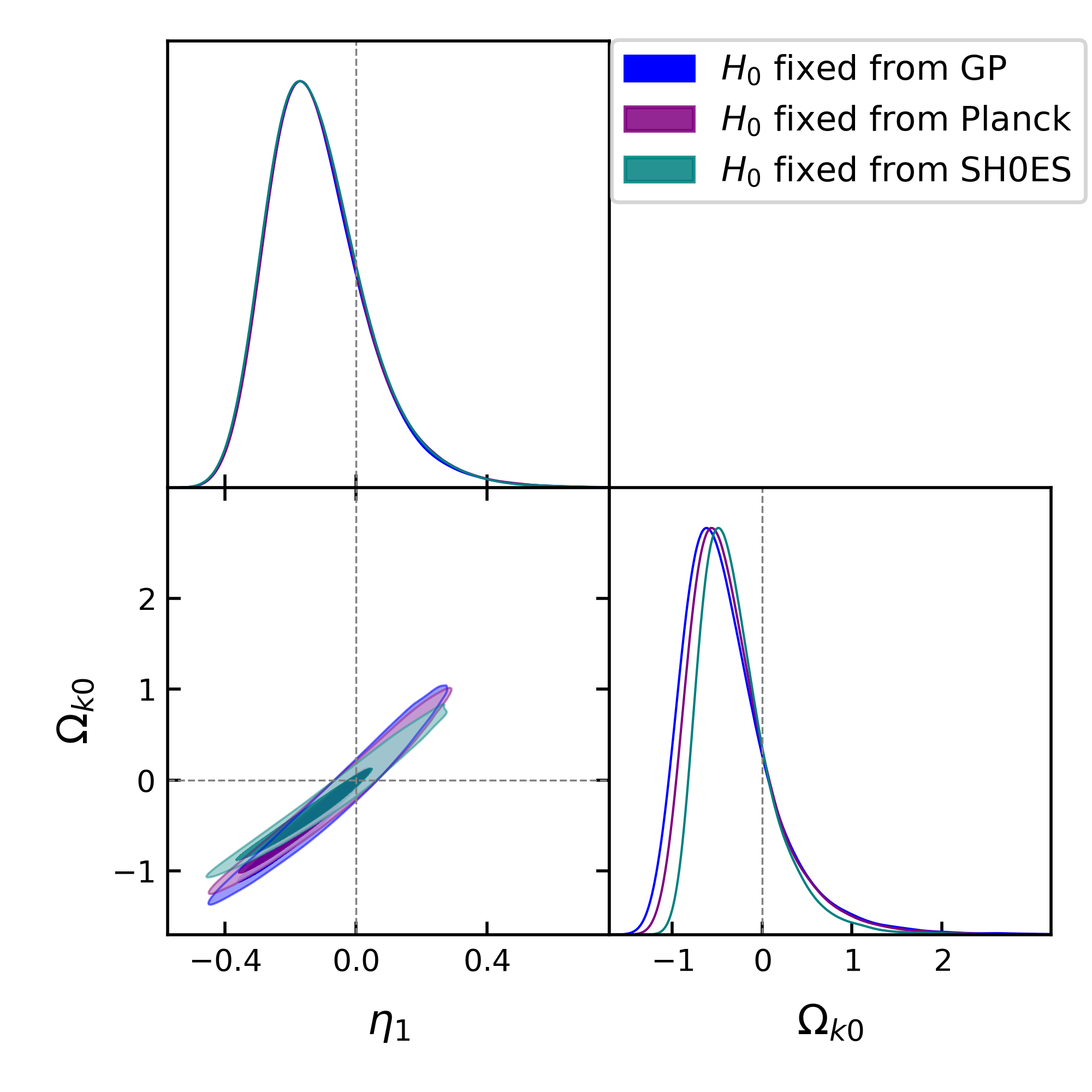}}
\caption{Posterior distributions and 68\% and 95\% confidence levels for $\eta_1$ and $\Omega_{k0}$ corresponding to the parameterizations P1–P4. The results use DESI BAO measurements with different $H_0$ priors (GP, Planck, SH0ES), in direct correspondence with Table~\ref{tab_cddr_bao_desi_results}. The dashed vertical lines indicate $\eta_1 = 0$ and $\Omega_{k0} = 0$, representing the standard cosmic distance duality relation and a spatially flat universe, respectively, to highlight potential deviations from these standard values.}
\label{fig_cddr_bao_desi_contours}
\end{figure*}

{We perform a joint test of the CDDR and spatial curvature in which $d_L(z)$ is obtained from a Gaussian Process reconstruction of $H(z)$ and $d_A(z)$ is provided by BAO measurements. Our procedure follows a clear roadmap: first, we reconstruct $H(z)$ from cosmic chronometer data to compute the $d_L(z)$, considering three different $H_0$ obtained from the GP, Planck, and SH0ES priors to assess their impact. Second, we consider three BAO datasets to obtain $d_A(z)$: $(i)$ the 2D BAO measurement including 15 data points of $\theta_{\rm BAO}$, $(ii)$ the 3D BAO measurement from DES~Y6 and BOSS/eBOSS including 5 data points of $d_{A}(z)/r_d$, and $(iii)$ the latest 3D DESI DR2 dataset consisting of 6 data points of $d_{A}(z)/r_d$. We explicitly check for systematic consistency between these angular and anisotropic BAO datasets. Third, we combine $d_L(z)$ and $d_A(z)$ through the distance duality relation parameter $\eta(z)$, adopting four representative parameterizations to study the redshift evolution of CDDR.}
A summary of the constraint results is given in Tables~\ref{tab_cddr_bao_2d_results}–\ref{tab_cddr_bao_desi_results} and illustrated in Figures~\ref{fig_cddr_bao_2d_contours}–\ref{fig_cddr_bao_desi_contours}.

\begin{table*}[ht]
\centering
\renewcommand{\arraystretch}{2}
\setlength{\tabcolsep}{4pt}
\begin{tabular}{ccccccc}
\hline
\multirow{2}{*}{Parameterization} 
& \multicolumn{3}{c}{$\eta_1$} 
& \multicolumn{3}{c}{$\Omega_{k0}$} \\
\cline{2-7}
& $H_0^{\mathrm{GP}}$ & $H_0^\mathrm{Planck}$ & $H_0^\mathrm{SH0ES}$ 
& $H_0^\mathrm{GP}$ & $H_0^\mathrm{Planck}$ & $H_0^\mathrm{SH0ES}$ \\
\hline
P1 & $0.072^{+0.106}_{-0.094}$ & $0.069^{+0.106}_{-0.092}$ & $0.072^{+0.108}_{-0.092}$ 
   & $1.061^{+0.925}_{-0.887}$ & $0.953^{+0.859}_{-0.799}$ & $0.834^{+0.740}_{-0.674}$ \\
\hline
P2 & $-0.277^{+0.108}_{-0.096}$ & $-0.278^{+0.106}_{-0.095}$ & $-0.274^{+0.108}_{-0.097}$ 
   & $-0.557^{+0.411}_{-0.373}$ & $-0.510^{+0.369}_{-0.342}$ & $-0.426^{+0.325}_{-0.292}$ \\
\hline
P3 & $-0.211^{+0.105}_{-0.096}$ & $-0.212^{+0.105}_{-0.094}$ & $-0.210^{+0.108}_{-0.095}$ 
   & $-0.815^{+0.643}_{-0.623}$ & $-0.753^{+0.599}_{-0.562}$ & $-0.629^{+0.512}_{-0.476}$ \\
\hline
P4 & $-0.249^{+0.110}_{-0.095}$ & $-0.247^{+0.107}_{-0.095}$ & $-0.243^{+0.108}_{-0.096}$ 
   & $-0.852^{+0.546}_{-0.461}$ & $-0.779^{+0.497}_{-0.421}$ & $-0.654^{+0.432}_{-0.358}$ \\
\hline
\end{tabular}
\caption{The best-fit values and its 68\% confidence level uncertainties for the parameters $\eta_1$ and $\Omega_{k0}$ inferred from 2D BAO measurements with different $H_0$ priors (GP, Planck, SH0ES). Here, $\eta_1 = 0$ corresponds to the standard cosmic distance duality relation, and $\Omega_{k0} = 0$ corresponds to a spatially flat universe. }
\label{tab_cddr_bao_2d_results}
\end{table*}

In the case of the 2D BAO measurement (see Figure~\ref{fig_cddr_bao_2d_contours} and Table~\ref{tab_cddr_bao_2d_results}), the constraints on $\eta_1$ and $\Omega_{k0}$ are consistent with $\eta_1=0$ and $\Omega_{k0}=0$ {within the 99\% confidence level}. 
{Therefore, we find no strong evidence for a violation of the CDDR or for a non-flat universe. {Under the P1 parameterization, the best-fit values of both $\eta_1$ and $\Omega_{k0}$ are positive, and the best-fit value of $\Omega_{k0}$ corresponds to an open geometry; however, a spatially flat universe remains fully accommodated within the 95\% confidence level. 
It is worth noting that the P1 form, $\eta(z)=1+\eta_1 z$, effectively behaves as a low-redshift expansion and may therefore be less reliable when extrapolated to higher redshifts, which offers a natural explanation for its differences relative to the P2--P4 parameterizations. 
The constraints show only minor variations across different $H_0$ priors (GP, Planck, SH0ES), but all are statistically consistent with $\Omega_{k0}=0$ and $\eta_1=0$ (see Table~\ref{tab_cddr_bao_2d_results} for full numerical values).}  
{In contrast, the P2–P4 parameterizations shift the best-fit values of both parameters to negative values, although the associated uncertainties remain large and the results stay consistent with $\Omega_{k0}=0$ at the 95\% confidence level.} 
This sign change shows that the inferred parameters depend on the specific CDDR parameterization. Moreover, when we change the $H_0$ prior, the best-fit value of $\Omega_{k0}$ shifts slightly and its constraints become somewhat tighter}, while the best-fit value of $\eta_1$ is essentially unchanged. In all cases, the results remain consistent with $\Omega_{k0}=0$ {at 95\% confidence level}, and the qualitative conclusions are unchanged. {The two-dimensional contours in Figure~\ref{fig_cddr_bao_2d_contours} show a strong positive degeneracy between $\eta_1$ and $\Omega_{k0}$ for all parameterizations and $H_0$ priors.}    
   

\begin{table*}[h!]
\centering
\renewcommand{\arraystretch}{2}
\begin{tabular}{ccccccc}
\hline
\multirow{2}{*}{Parameterization} & \multicolumn{3}{c}{$\eta_1$} & \multicolumn{3}{c}{$\Omega_{k0}$} \\
\cline{2-7}
 & $H_0^\mathrm{GP}$ & $H_0^\mathrm{Planck}$ & $H_0^\mathrm{SH0ES}$ & $H_0^\mathrm{GP}$ & $H_0^\mathrm{Planck}$ & $H_0^\mathrm{SH0ES}$ \\
\hline
P1 & $0.003^{+0.133}_{-0.095}$ & $0.000^{+0.134}_{-0.094}$ & $-0.003^{+0.145}_{-0.094}$ & $-0.017^{+1.023}_{-0.864}$ & $-0.032^{+0.944}_{-0.794}$ & $-0.052^{+0.862}_{-0.672}$ \\
\hline
P2 & $-0.281^{+0.172}_{-0.139}$ & $-0.281^{+0.173}_{-0.141}$ & $-0.282^{+0.170}_{-0.140}$ & $-0.608^{+0.379}_{-0.324}$ & $-0.559^{+0.351}_{-0.299}$ & $-0.475^{+0.292}_{-0.257}$ \\
\hline
P3 & $-0.268^{+0.159}_{-0.126}$ & $-0.267^{+0.155}_{-0.127}$ & $-0.263^{+0.150}_{-0.129}$ & $-1.204^{+0.736}_{-0.679}$ & $-1.103^{+0.659}_{-0.627}$ & $-0.921^{+0.543}_{-0.543}$ \\
\hline
P4 & $-0.256^{+0.157}_{-0.129}$ & $-0.260^{+0.154}_{-0.123}$ & $-0.259^{+0.157}_{-0.127}$ & $-0.936^{+0.540}_{-0.398}$ & $-0.877^{+0.488}_{-0.349}$ & $-0.740^{+0.420}_{-0.305}$ \\
\hline
\end{tabular}
\caption{The best-fit values and its 68\% confidence level uncertainties for the parameters $\eta_1$ and $\Omega_{k0}$ inferred from 3D BAO measurements with different $H_0$ priors (GP, Planck, SH0ES). Here, $\eta_1 = 0$ corresponds to the standard cosmic distance duality relation, and $\Omega_{k0} = 0$ corresponds to a spatially flat universe. }
\label{tab_cddr_bao_3d_results}
\end{table*}

Using the 3D BAO measurements, we find that the constraints on $\eta_1$ and $\Omega_{k0}$ remain consistent with $\eta_1=0$ and $\Omega_{k0}=0$ within the 95\% confidence level (see Figure~\ref{fig_cddr_bao_3d_contours} and Table~\ref{tab_cddr_bao_3d_results}). {These results provide no strong support for a CDDR violation or for non-flat spatial curvature across the four parameterizations.} 
A {significant} $\eta_1$–$\Omega_{k0}$ degeneracy is again observed, {as} seen in Figure~\ref{fig_cddr_bao_3d_contours}, which is similar to the 2D BAO case. 
{Under the P1 parameterization, the transition from 2D to 3D BAO measurements leads to a sign change in the best-fit values of $\eta_1$ and $\Omega_{k0}$. Nevertheless, the corresponding shifts are modest, remaining within the quoted $1\sigma$ uncertainties and at the level of a few $\mathcal{O}( 10^{-2})$ to $\mathcal{O}(10^{-1})$, with the 3D BAO constraints placing both parameters very close to zero. This behavior indicates that, for this particular parameterization, the inferred values of $\eta_1$ and $\Omega_{k0}$ are sensitive not only to the functional form of $\eta(z)$ but also to the adopted dataset. 
Such sign changes should not be interpreted as a physical reversal; rather, they reflect the fact that the P1 form, $\eta(z)=1+\eta_1 z$, effectively corresponds to a low-redshift expansion and can therefore become more sensitive to data selection when applied over a wide redshift range. 
Meanwhile, we find that $\eta_1$ is comparatively insensitive to the choice of the $H_0$ prior, whereas $\Omega_{k0}$ exhibits a mild dependence, with its best-fit value moving slightly closer to zero and its uncertainty modestly decreasing for higher $H_0$ priors. 
These effects do not alter our main conclusion that the results remain consistent with a spatially flat universe.}

\begin{table*}[h!]
\centering
\renewcommand{\arraystretch}{2}
\begin{tabular}{ccccccc}
\hline
\multirow{2}{*}{Parameterization} & \multicolumn{3}{c}{$\eta_1$} & \multicolumn{3}{c}{$\Omega_{k0}$} \\
\cline{2-7}
 & $H_0^\mathrm{GP}$ & $H_0^\mathrm{Planck}$ & $H_0^\mathrm{SH0ES}$ & $H_0^\mathrm{GP}$ & $H_0^\mathrm{Planck}$ & $H_0^\mathrm{SH0ES}$ \\
\hline
P1 & $-0.071^{+0.133}_{-0.076}$ & $-0.071^{+0.131}_{-0.079}$ & $-0.065^{+0.146}_{-0.079}$ & $-0.624^{+1.086}_{-0.783}$ & $-0.581^{+0.987}_{-0.743}$ & $-0.444^{+0.911}_{-0.624}$ \\
\hline
P2 & $-0.125^{+0.222}_{-0.163}$ & $-0.126^{+0.212}_{-0.164}$ & $-0.131^{+0.217}_{-0.159}$ & $-0.222^{+0.372}_{-0.305}$ & $-0.207^{+0.335}_{-0.281}$ & $-0.180^{+0.289}_{-0.237}$ \\
\hline
P3 & $-0.136^{+0.172}_{-0.132}$ & $-0.149^{+0.166}_{-0.126}$ & $-0.153^{+0.171}_{-0.121}$ & $-0.522^{+0.638}_{-0.566}$ & $-0.532^{+0.581}_{-0.501}$ & $-0.462^{+0.504}_{-0.409}$ \\
\hline
P4 & $-0.137^{+0.163}_{-0.125}$ & $-0.135^{+0.164}_{-0.125}$ & $-0.135^{+0.164}_{-0.128}$ & $-0.471^{+0.552}_{-0.374}$ & $-0.425^{+0.518}_{-0.349}$ & $-0.363^{+0.442}_{-0.302}$ \\
\hline
\end{tabular}
\caption{The best-fit values and its 68\% confidence level uncertainties for the parameters $\eta_1$ and $\Omega_{k0}$ inferred from 3D DESI BAO measurements with different $H_0$ priors (GP, Planck, SH0ES). Here, $\eta_1 = 0$ corresponds to the standard cosmic distance duality relation, and $\Omega_{k0} = 0$ corresponds to a spatially flat universe.}
\label{tab_cddr_bao_desi_results}
\end{table*}

We find that the constraints from the 3D DESI BAO measurements on $\eta_1$ and $\Omega_{k0}$ are consistent with $\eta_1=0$ and $\Omega_{k0}=0$ at the 68\% confidence level for all four parameterizations, as shown in Figure~\ref{fig_cddr_bao_desi_contours} and Table~\ref{tab_cddr_bao_desi_results}. 
These results provide additional support for the validity of the CDDR and compatibility with a spatially flat universe. 
As in the other cases, $\eta_1$ and $\Omega_{k0}$ exhibit a clear positive correlation. 
{Consistent with the case of 3D BAO measurement, we find that using the 3D DESI BAO dataset also leads to modest shifts in the inferred parameters $\eta_1$ and $\Omega_{k0}$. 
In particular, the P1 parameterization with the 3D DESI BAO data provides slightly more negative best-fit values for both $\eta_1$ and $\Omega_{k0}$, while the exact numerical results for different $H_0$ priors are listed in Table~\ref{tab_cddr_bao_desi_results}. 
These shifts further illustrate that the choice of BAO dataset can affect the best-fit values of $\eta_1$ and $\Omega_{k0}$; however, all results remain statistically consistent with zero.}
{The $H_0$ prior has a weak impact on $\eta_1$ but a non-negligible impact on $\Omega_{k0}$, consistent with the 2D and 3D BAO analyses.} It is found that a higher $H_0$ prior move the best-fit value of $\Omega_{k0}$ towards zero and slightly reduces its uncertainty. 
Finally, although recent 3D DESI BAO analyses report a mild preference for dynamical dark energy relative to the $\Lambda$CDM model, our results show no evidence for the deviations from $\eta_1=0$.

\begin{figure*}[htbp]
    \centering
    \subfigure[$H_0^\mathrm{GP}$ prior]{
        \includegraphics[width=0.3\textwidth]{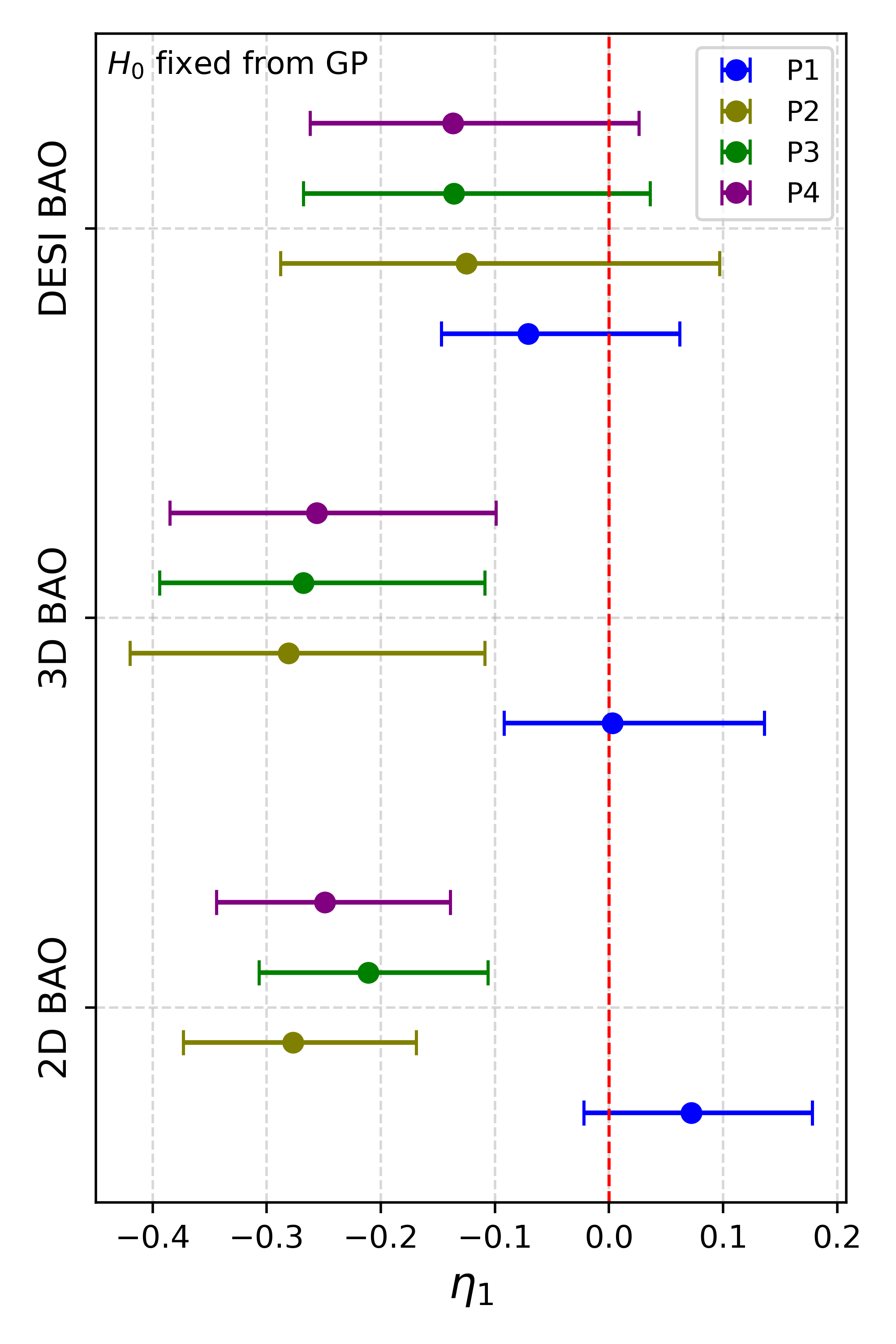}
        \label{fig_eta1_gp}
    }\hfill
    \subfigure[$H_0^\mathrm{Planck}$ prior]{
        \includegraphics[width=0.3\textwidth]{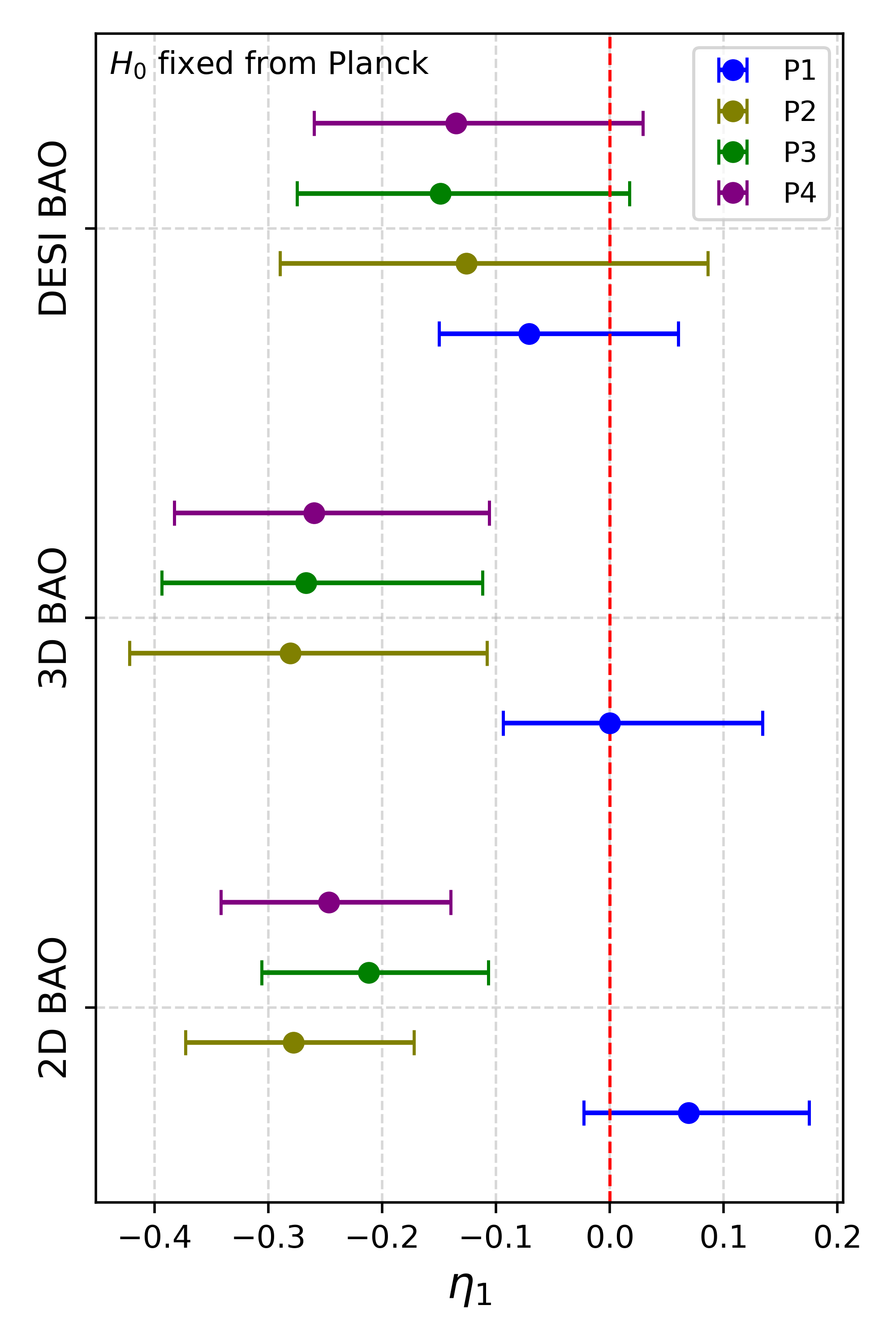}
        \label{fig_eta1_planck}
    }\hfill
    \subfigure[$H_0^\mathrm{SH0ES}$ prior]{
        \includegraphics[width=0.3\textwidth]{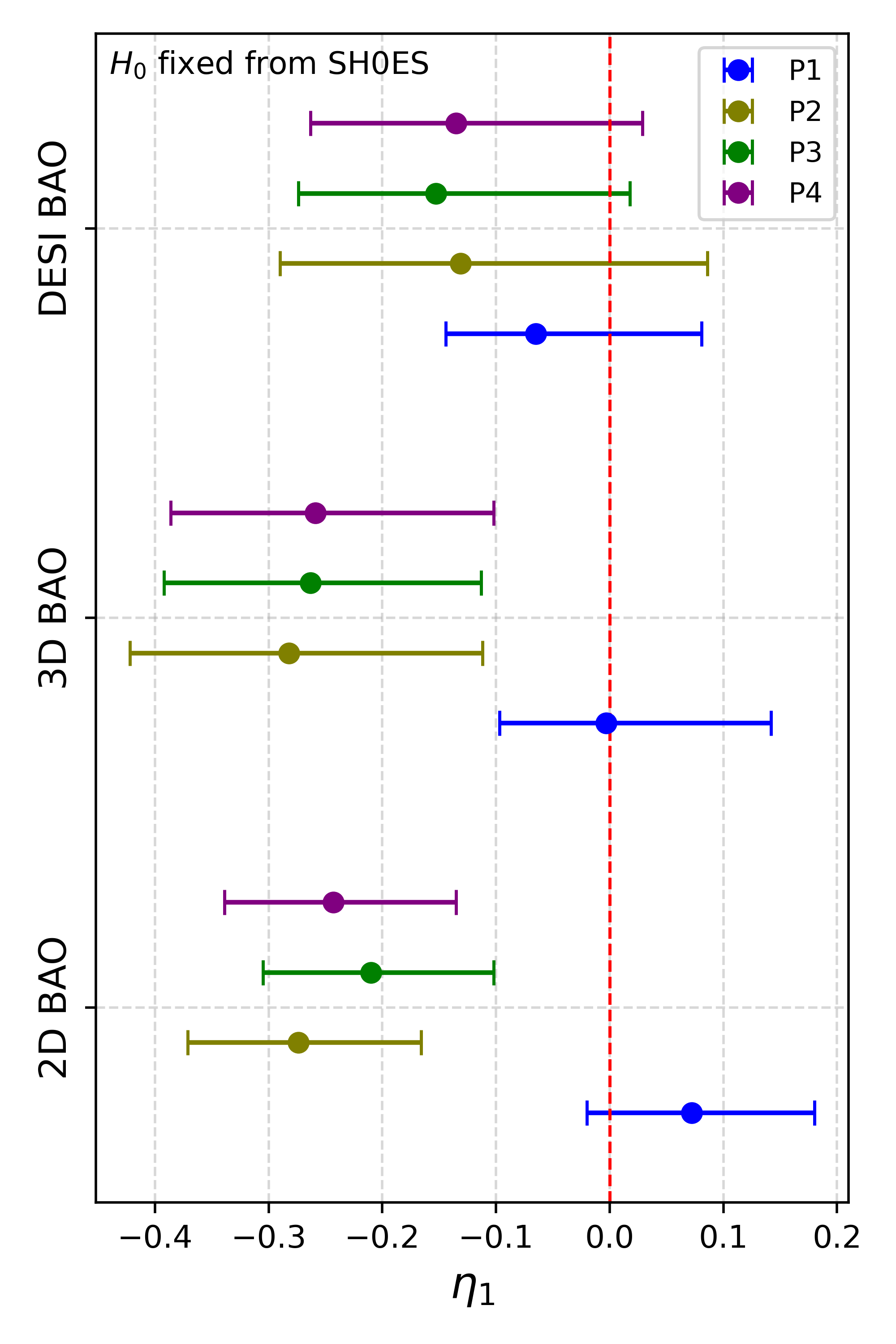}
        \label{fig_eta1_shoes}
    }
    \caption{The whisker plots show the best-fit values and 68\% confidence intervals for $\eta_1$ under three $H_0$ priors: GP, Planck, and SH0ES. Each horizontal row on the y-axis corresponds to a BAO dataset, labeled as 2D BAO, 3D BAO, and 3D DESI DR2. Different colors indicate the four parameterizations P1–P4, and the red dashed vertical line at $\eta_1 = 0$ represents the standard cosmic distance duality relation. }
    \label{fig_eta1_whiskers}
\end{figure*}

\begin{figure*}[htbp]
    \centering
    \subfigure[$H_0^\mathrm{GP}$ prior]{
        \includegraphics[width=0.3\textwidth]{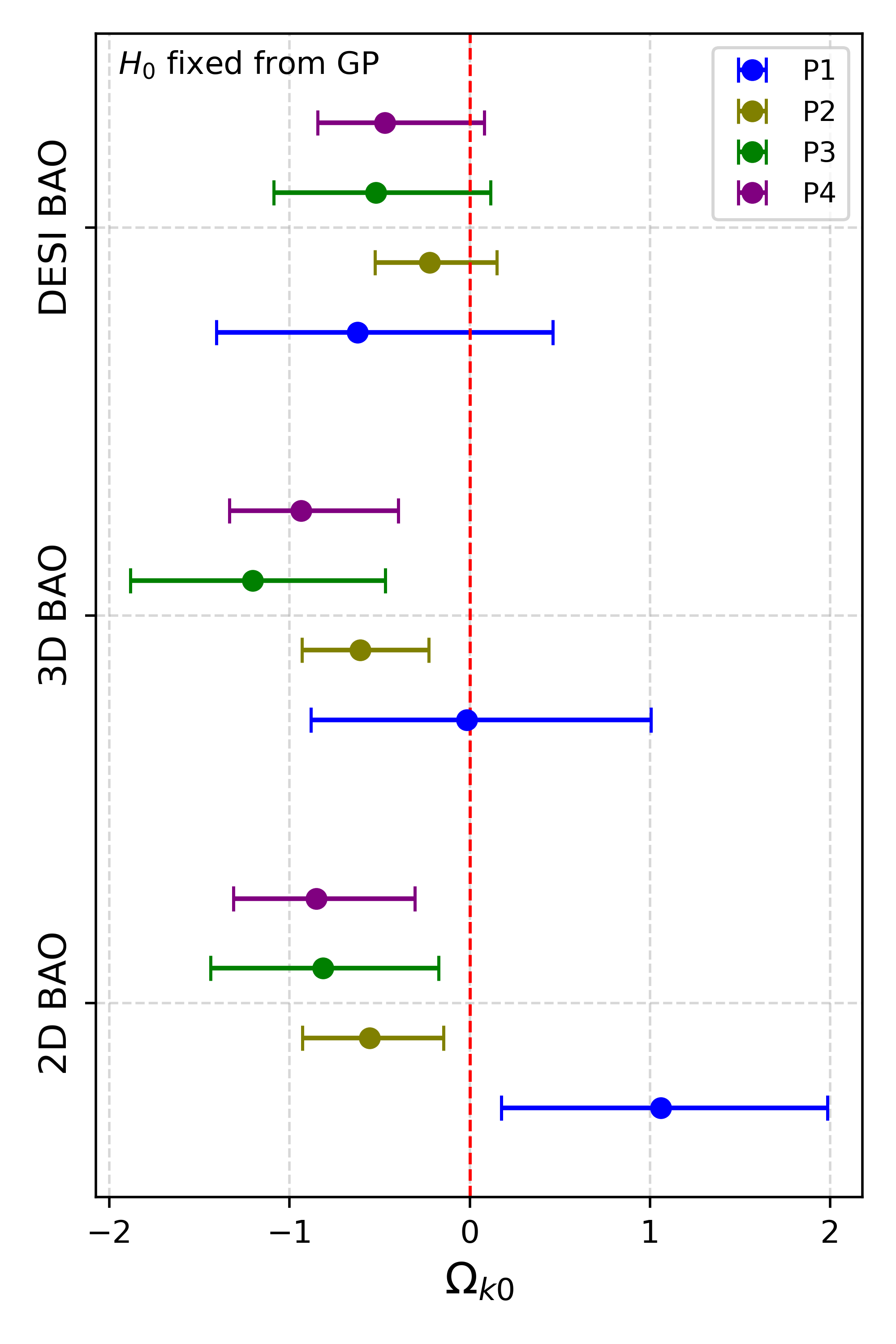}
        \label{fig_ok_gp}
    }\hfill
    \subfigure[$H_0^\mathrm{Planck}$ prior]{
        \includegraphics[width=0.3\textwidth]{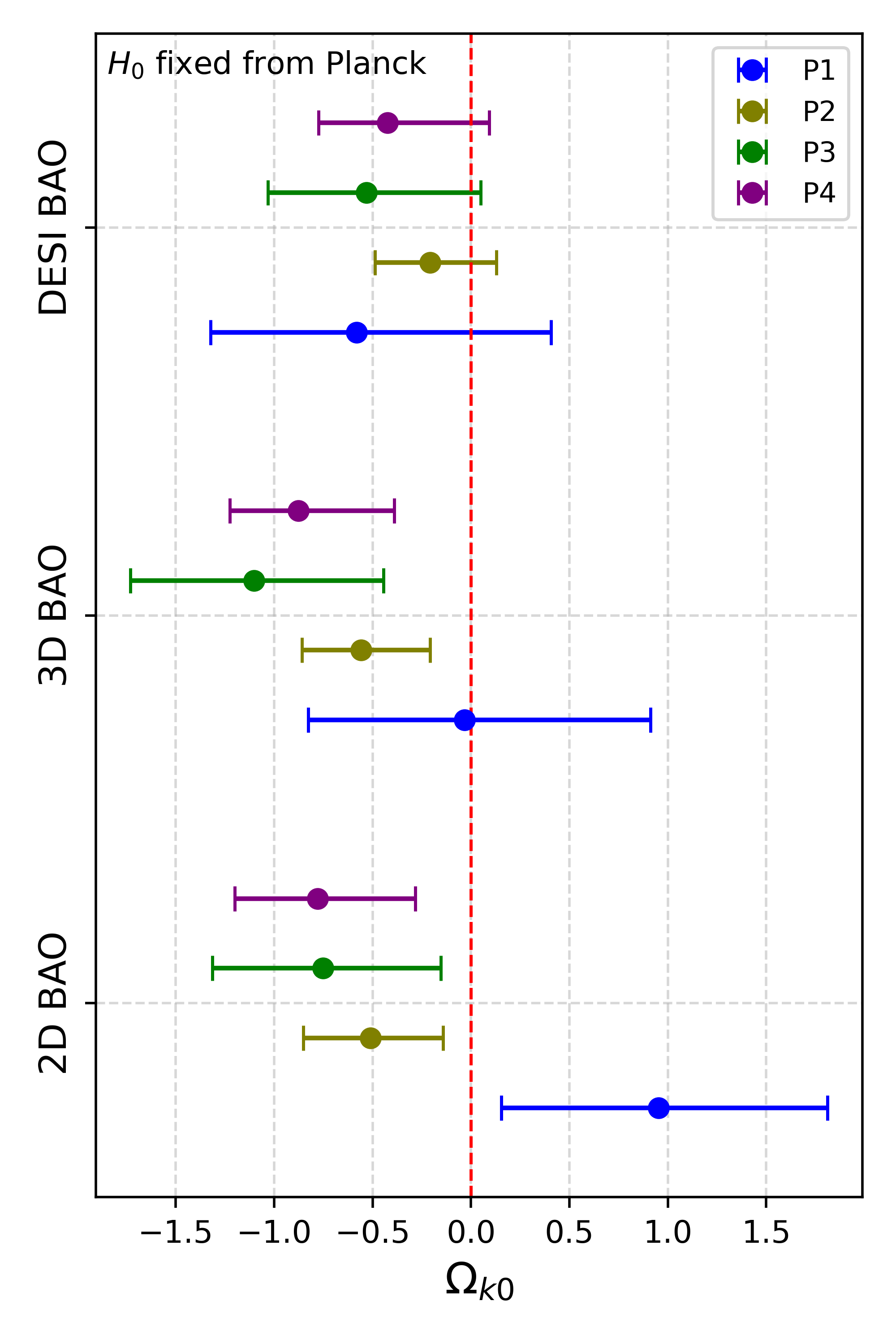}
        \label{fig_ok_planck}
    }\hfill
    \subfigure[$H_0^\mathrm{SH0ES}$ prior]{
        \includegraphics[width=0.3\textwidth]{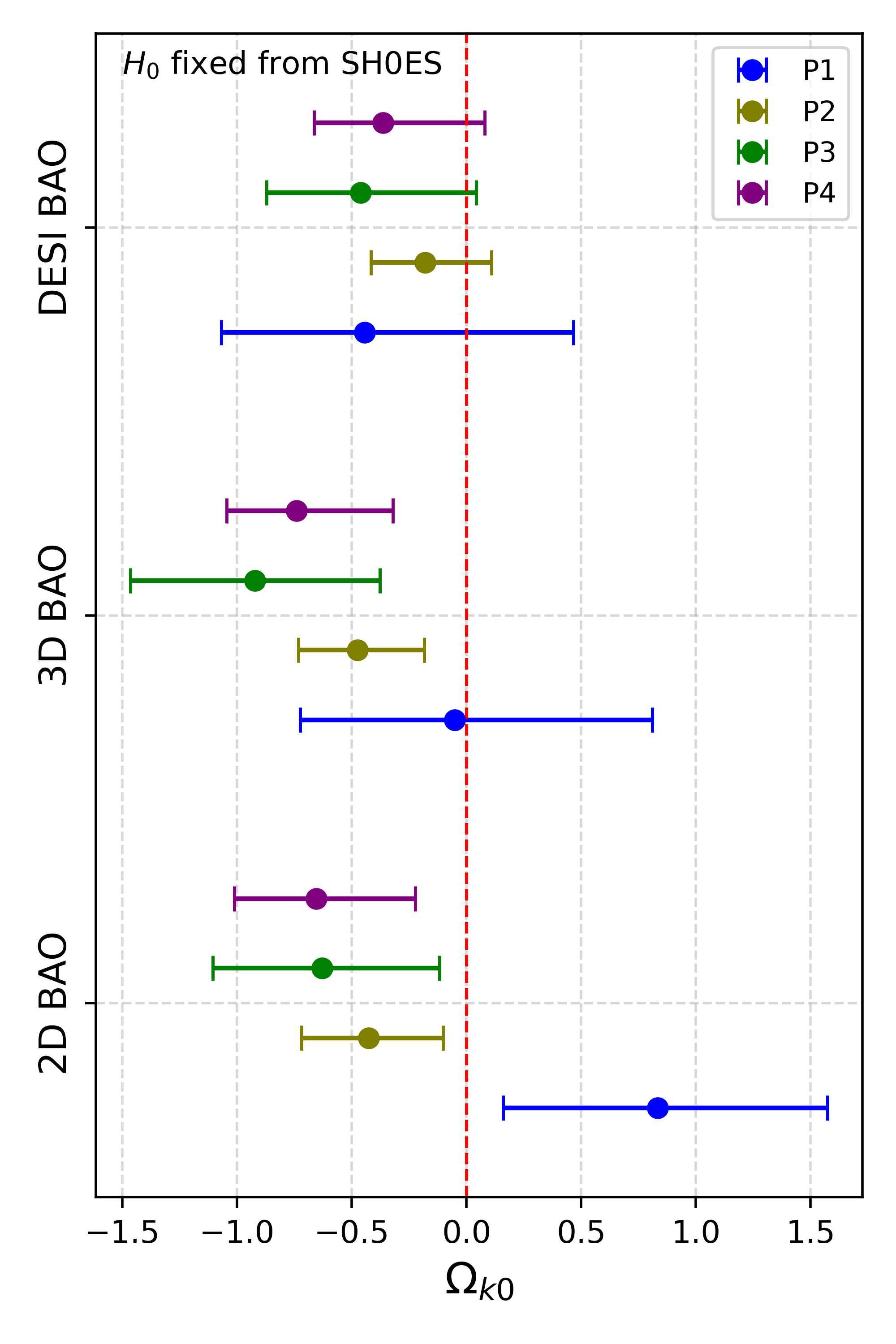}
        \label{fig_ok_shoes}    
    }
    \caption{The whisker plots show the best-fit values and 68\% confidence intervals for $\Omega_{k0}$ under three $H_0$ priors: GP, Planck, and SH0ES. Each horizontal row on the y-axis corresponds to a BAO dataset, labeled as 2D BAO, 3D BAO, and 3D DESI DR2. Different colors indicate the four parameterizations P1–P4, and the red dashed vertical line at $\Omega_{k0}=0$ represents a spatially flat universe. }
    
    \label{fig_ok_whiskers}  
\end{figure*}

{For all BAO measurements we adopted, the posteriors of $\eta_1$ are centered around zero and those of $\Omega_{k0}$ include zero within the 95\% confidence level, indicating no evidence for a CDDR violation or non-flat curvature.}
In Figure~\ref{fig_eta1_whiskers} and Figure~\ref{fig_ok_whiskers}, it displays that the best-fit values of $\eta_1$ and $\Omega_{k0}$, as well as their intervals. Although the errors are broader, the intervals largely overlap. 
Therefore, the constraints from 2D BAO, 3D BAO, and 3D DESI BAO are consistent with each other, suggesting that the potential tension between 2D and 3D BAO measurements does not materially affect the test of CDDR or the inference of $\Omega_{k0}$.
{In all cases of BAO datasets, the best-fit values of $\Omega_{k0}$ may favor a non-flat universe, but $\Omega_{k0}=0$ is still contained within the 95\% confidence level.}
Among these BAO measurements, the 3D BAO measurements prefer a more negative $\Omega_{k0}$. 
As the most up-to-date and precise BAO dataset currently available, the 3D DESI BAO prefers a flat universe and the validity of the CDDR.

For the different CDDR parameterizations, the P2, P3, and P4 models give very similar results with various BAO datasets. 
By contrast, the P1 model shows a small deviation from the others, which may be due to its linear form, $\eta(z)=1+\eta_1 z$, that can grow without bound at high redshift.
From the above results, it indicates that changing the $H_0$ prior has little effect on $\eta_1$. 
In contrast, it shows a possible positive correlation between $H_0$ and $\Omega_{k0}$. The constraints on $\Omega_{k0}$ depend on the $H_0$ prior; moving from a lower $H_0$ to a higher $H_0$ shifts $\Omega_{k0}$ toward zero and slightly tightens the error bar. 
However, the correlation is modest and does not change our qualitative conclusions.

{For comparison, \citep{2022JCAP...01..053K} estimated $\eta_1$ and $\Omega_{k0}$ using SNe Ia and $H(z)$ datasets, which required a prior on the absolute magnitude $M_B$. In contrast, our analysis combines $H(z)$ data with both 2D and 3D BAO measurements, including the latest DESI DR2 dataset. This approach avoids the $M_B$ prior and extends the redshift coverage. To our knowledge, this study presents the first systematic comparison of different types of BAO measurements within a unified framework to assess their impact on the validity of the CDDR and $\Omega_{k0}$. Recently, \citep{Alfano:2025gie} tested four parameterizations of the CDDR in both model-independent and model-dependent frameworks using $H(z)$, galaxy cluster data, SNe Ia, and BAO observations. They found no evidence for a violation of the CDDR, although a slight spatial curvature cannot be entirely excluded, which is consistent with our results. Quantitatively, the typical uncertainties obtained in this work are $\sigma(\eta_1)\sim\mathcal{O}(10^{-1})$ and $\sigma(\Omega_{k0})\sim\mathcal{O}(10^{-1})$, which are comparable to those reported in previous joint analyses based on other cosmological probes. 
We further find that the variations in the best-fit values of $\eta_1$ and $\Omega_{k0}$ are mainly driven by the choice of the $\eta(z)$ parameterization and the adopted $H_0$ prior. In contrast, the uncertainties on $\eta_1$ and $\Omega_{k0}$ are largely determined by the statistical precision of the BAO measurements.
In addition, a broad range of independent studies focusing exclusively on tests of the CDDR, without simultaneously constraining on $\Omega_{k0}$, have consistently found no significant deviation from $\eta(z)=1$, in agreement with our results \citep{AS7,UV1,2025ApJ...979....2Q,Zheng:2025cgq,Kanodia:2025jqh}.}


\section{Conclusions}\label{sec_disc_conc}

We present the first comprehensive analysis that simultaneously tests $\eta(z)$ and $\Omega_{k0}$, using three types of BAO measurements for $d_A(z)$ and Gaussian Process reconstructions of Cosmic Chronometer $H(z)$ to obtain $d_L(z)$. Specifically, we systematically compare different BAO measurements, 2D BAO, 3D BAO, and 3D DESI BAO, to assess whether the existing tension between the angular and anisotropic BAO data would affect the estimations of $\eta(z)$ and $\Omega_{k0}$. Furthermore, to comprehensively test the CDDR and examine the degeneracy between $\eta(z)$ and $\Omega_{k0}$, we adopt four representative parameterizations of $\eta(z)$. In addition, we quantify how different $H_0$ priors affect the constraints on $\eta_1$ and $\Omega_{k0}$.

\vspace{1.5mm}
Our main conclusions are as follows:

\begin{itemize}
    \item {Our results are consistent with $\eta(z)=1$ and $\Omega_{k0}=0$ at 99\% confidence level and 95\% confidence level, respectively. They support the validity of the CDDR and the compatibility with a flat universe.} {Although the best-fit value of $\Omega_{k0}$ deviates slightly from zero, this deviation remains well within the statistical uncertainties and does not exceed the 95\% confidence level. Consequently, any indication of a non-flat geometry is not statistically significant. It is noteworthy that small departures from the standard CDDR or from exact spatial flatness cannot be ruled out with the current precision of the data.} 
    In addition, the 2D posterior contours in the $(\eta_1,\Omega_{k0})$ plane show a clear positive correlation. It points to why different BAO datasets or $H_0$ priors could shift the best-fit value slightly without producing a significant change in the overall consistency with $\eta_1=0$ and $\Omega_{k0}=0$.         

    \item Although the best-fit values from different BAO datasets show a small inconsistency, their 68\% credible intervals largely overlap. Therefore, the prospective tension between 2D and 3D BAO measurements does not strongly affect the estimations of $\eta_1$ and $\Omega_{k0}$. Further progress will require higher-precision BAO datasets to investigate this issue more decisively.

    \item Through careful analysis, we find that the choice of the $H_0$ prior does not have a significant impact on the correlations among parameters.  For $\eta_1$, the effect is weak across all datasets and parameterizations. For $\Omega_{k0}$, however, moving from a lower to a higher $H_0$ prior shifts the best-fit value toward zero and slightly tightens the 68\% confidence level. This prior sensitivity is modest, since most error bars still overlap, and it does not change our qualitative conclusions.
\end{itemize}
In conclusion, our analysis provides support for the validity of the cosmic distance duality relation and a flat universe. {With the current data, the constraints on $\eta_1$ and $\Omega_{k0}$ are dominated by statistical uncertainties, so differences between 2D and 3D BAO measurements remain subdominant, although they could become more relevant as BAO and Cosmic Chronometer datasets improve.} Future work could benefit from larger datasets, higher redshifts, and improved precision, such as the Euclid Space Telescope \citep{2012SPIE.8442E..0ZA,2020A&A...644A..80M}. In parallel, using multiple cosmological probes and exploring alternative parameterizations could help further clarify the relationship between $\Omega_{k0}$ and $\eta(z)$ and its implications for fundamental physics.


\section*{Acknowledgments}
We thank the anonymous referee for helpful comments that improved this work. Kumar, D. is supported by the Startup Research Fund of the Henan Academy of Sciences under Grant number 241841219.             
Zheng, J. is supported by the National Natural Science Foundation of China under Grant No. 12403002 and 12433001; The Scientific and Technological Research Project of Henan Academy of Science (Project No. 20252345001); The Startup Research Fund of Henan Academy of Sciences (Project No. 241841221). 
You, Z.-Q. is supported by the National Natural Science Foundation of China under Grant No. 12305059 and 12433001; The Startup Research Fund of Henan Academy of Sciences (Project No. 241841224); The Scientific and Technological Research Project of Henan Academy of Science (Project No. 20252345003); Joint Fund of Henan Province Science and Technology R\&D Program (Project No. 235200810111); Henan Province High-Level Talent Internationalization Cultivation (Project No. 2024032).
Qiang, D.-C. is supported by the National Natural Science Foundation of China under grant No. 12505070, the Henan Provincial Natural Science Foundation No. 252300420902, and the Startup Research Fund of Henan Academy of Sciences No.241841222.

\bibliography{main}{}
\bibliographystyle{aasjournal}

\end{document}